\documentclass[reprint,prl,aps]{revtex4-2}
\usepackage{graphicx}
\usepackage{color}
\usepackage{dcolumn}
\usepackage{bm}
\usepackage{amssymb}
\usepackage{color,amsmath}
\usepackage{mathrsfs}
\usepackage{soul,xcolor} 
\setstcolor{red} 
\usepackage{epstopdf}
\usepackage{gensymb}
\usepackage[hypertexnames=false]{hyperref}
\hypersetup{hidelinks}
\usepackage{xurl}
\usepackage{siunitx}

\usepackage{soul}

\begin{document}

\title{Direct synthesis of moir\'{e} graphene with twist-tuned hybridization}

\author{Seunghyun Park$^{1}$}
\author{Daniel T. Larson$^{1}$}
\author{Zeyu Hao$^{2}$}
\author{Takashi Taniguchi$^{3}$}
\author{Kenji Watanabe$^{3}$}
\author{Efthimios Kaxiras$^{1}$}
\author{Yuan Cao$^{2}$}
\author{Amir Yacoby$^{1}$}

\affiliation{$^{1}$Department of Physics, Harvard University, Cambridge MA, 02138, USA}
\affiliation{$^{2}$Department of Physics, University of California at Berkeley, Berkeley, California 94720, USA}
\affiliation{$^{3}$International Center for Materials Nanoarchitectonics, National Institute for Materials Science, 1-1 Namiki, Tsukuba, Japan}
\maketitle

\addcontentsline{toc}{section}{Main Text}

\textbf{
Twisted van der Waals (vdW) systems provide a powerful platform to explore emergent correlated phenomena, including superconductivity \cite{yuanSC,cdean_wse2}, magnetism \cite{mote2xiaodong}, topological phases \cite{tBGSET,mote2hjpark,mote2nanosquid}, and beyond. Introducing the twist angle as a continuous tuning parameter greatly expands the accessible phase space by enabling controlled modification of the underlying Hamiltonian. However, precise and reproducible control of the twist angle remains a major challenge, hindering systematic exploration of the full phase diagram. Existing approaches partially address this bottleneck but are constrained by single-use device architectures, limited angular precision, or vertical probing geometries \cite{twistableelec,programmable,STMfolding,stminsitu,rofggcontact,CarozoR,qtm,roboticstack,oxygencvd,heterocvd,cvdjiwoongpark}. \\
Here we introduce a reconfigurable graphene-based moir\'{e} platform in which two independently fabricated layers are mechanically assembled and rotated in situ using a micro-electromechanical system (MEMS). We demonstrate the formation and robustness of the twist-induced moir\'{e} potential using Raman spectroscopy, and establish continuous and deterministic control of the twist angle. The moir\'{e} potential folds the electronic band structure and activates the longitudinal optical (LO) phonon mode, enabling precise mapping of its dispersion and direct visualization of the moir\'{e} momentum. The angle-dependent electronic band hybridization is resolved through the evolution of the double-resonant 2D mode. Near \(0^\circ\), an emergent asymmetry of the $\Gamma$-point (G-mode) phonon reveals phonon band folding and hybridization driven by the moir\'{e} superlattice potential and interlayer relaxation. Our platform provides a reproducible, comprehensive Raman reference for twisted bilayer graphene, eliminating device-to-device variability. More broadly, it establishes the twist angle as a parameter that can be controlled deterministically, opening a pathway to systematic and reproducible studies of correlated and topological phenomena in moir\'{e} quantum materials. 
}

Moir\'{e} systems have emerged as a powerful platform for tailoring electronic band structures and electron-electron interactions, giving rise to a wealth of emergent correlated phenomena \cite{yuanSC,tBGCI,tBBG,KinFaiJieperspec,zuipaper,mote2xiaodong,mote2kinfai}. Because these electronic properties—particularly in graphene-based devices—are exquisitely sensitive to minute angular variations \cite{BMmodel,devereaux,miniarpestBG}, precise control over the twist angle is essential for designing the energy landscape and accurately probing the underlying physics. However, achieving this level of deterministic control remains a major experimental bottleneck. Previous efforts to fabricate moir\'{e} superlattices have explored numerous approaches, ranging from high-throughput assembly of device ensembles \cite{roboticstack,oxygencvd,heterocvd,cvdjiwoongpark,cdean_wse2,mote2arpes} to in situ manipulation using scanning probe tips \cite{twistableelec,programmable,STMfolding, stminsitu, rofggcontact,CarozoR} and dynamic polymer stamps \cite{stamptwist}. These techniques allow for the efficient, systematic production of diverse moir\'{e} structures. However, the ability to continuously and precisely tune the twist angle within a single integrated device remains an outstanding challenge.\\

Recently, pioneering experiments using the quantum twisting microscope (QTM) \cite{qtm, qtmphonon} have successfully demonstrated controllable twisting of graphene on a tip. These studies used electron tunneling between twisted graphene layers to probe electronic band structures \cite{qtm} and phonon dispersion \cite{qtmphonon}. The contact between the two graphene layers is sufficiently pristine to support coherent electron tunneling, and a signature of band hybridization was also reported \cite{qtmphonon}. 

Because these studies primarily aimed to extract electronic and vibrational information through tunneling spectroscopy, the measurements were performed in a regime where the tunneling signal could be interpreted within established theoretical frameworks \cite{qtmphonon}. 
Although these studies established a powerful platform for controllable twisting, a systematic investigation of interlayer band hybridization under active, continuous twist-angle control has yet to be performed and remains an important next step.

\begin{figure*}[ht]
\includegraphics[width=6in]{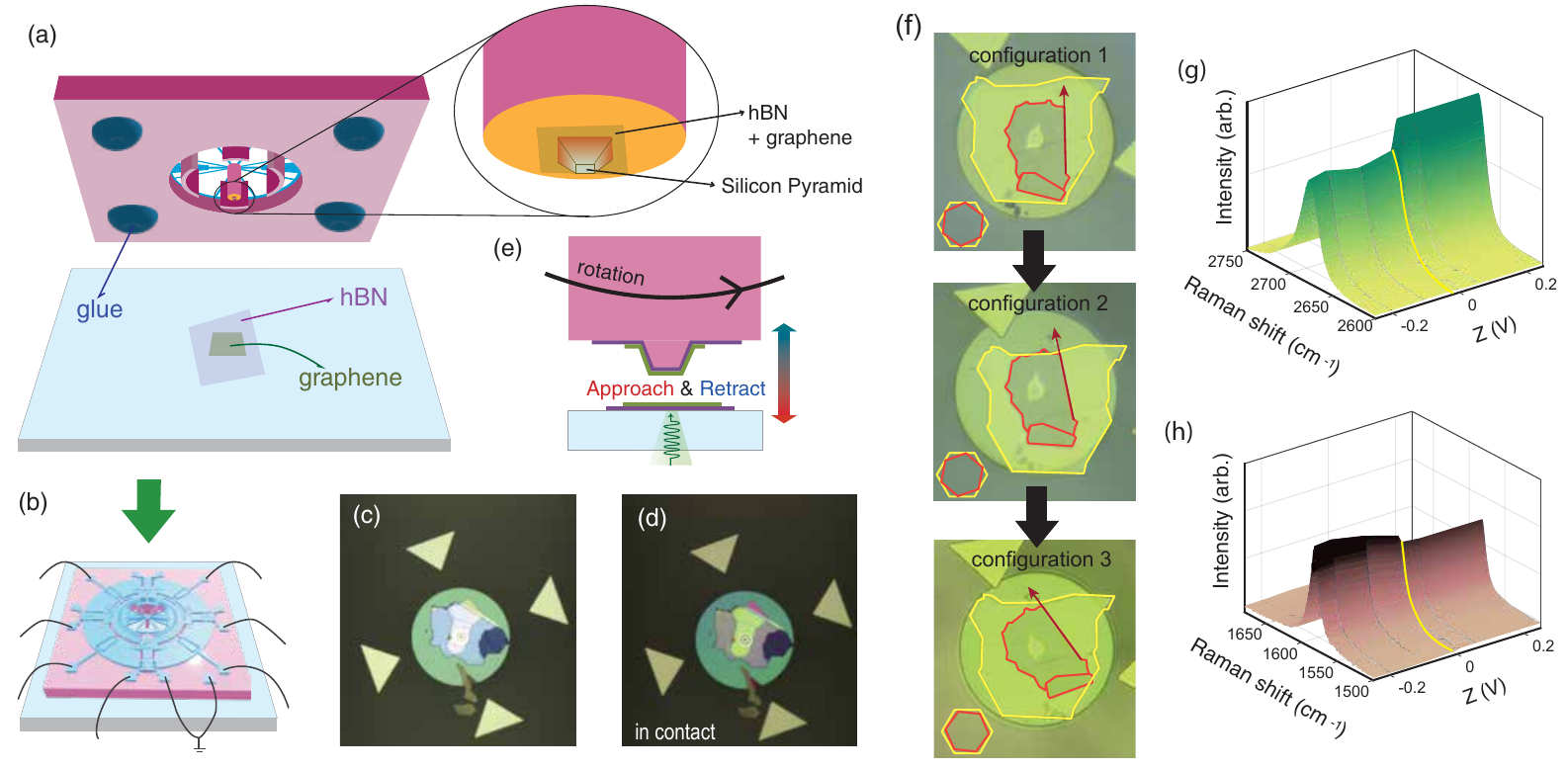} 
\centering
\caption{\textbf{Schematics of MEMS operation and optical microscopy images of MEMS devices}
\textbf{(a)} Schematic of the MEMS device with the pyramid facing the substrate. Glue droplets are deposited on the MEMS before the two pieces are combined. The hBN/graphene heterostructure is placed on the silicon pyramid with the graphene exposed toward the substrate side, allowing direct contact with the graphene on the substrate. 
\textbf{(b)} The assembled MEMS–substrate device is then wire-bonded to the PCB, enabling control of the angle and distance between the MEMS and the substrate.
\textbf{(c,d)} Optical microscope images showing (c) the MEMS in close proximity to the substrate without contact (approximately 1 \SI{}{\micro\meter} apart) and (d) the MEMS in contact with the substrate. The circular silicon island (the cross section of the rod) changes color as the MEMS approaches the substrate due to optical interference \cite{SIline}.
\textbf{(e)} Degrees of freedom of MEMS operation. The MEMS device enables in situ rotation of up to $\pm 10^\circ$ and vertical (Z) displacement of approximately \SI{2}{\micro\meter}. Optical spectroscopy is performed through the transparent fused silica substrate.
\textbf{(f)} Optical microscope images of three MEMS–substrate configurations after manual rotation. The red outline indicates the stack on the MEMS, while the yellow outline indicates the stack on the substrate.
\textbf{(g,h)} Raman G- and 2D-mode spectra measured while engaging the MEMS with the substrate. Positive Z corresponds to the MEMS being disengaged from the substrate, whereas negative Z corresponds to the MEMS pyramid pressing against the substrate under an applied voltage. The yellow line indicates the engagement position. 
}
\label{fig:1}
\end{figure*}

Here, we demonstrate reproducible in situ tuning of a twisted bilayer graphene (tBG) system using a MEMS device, named `MEGA2D' in its foundational study \cite{yuan2014onchip}. The constructed tBG system enables precise and controllable tuning of the twist angle between the two graphene layers. The MEMS has the same design as described in \cite{yuan2014onchip}. A silicon rod at the center of the structure is connected to rotating gears by thin beams and terminates in a $500$-nm-tall pyramid with a flat, $3.5\times3.5$ \SI{}{\micro\meter^2} top (Fig. \ref{fig:1}a). The pyramid is the only point that contacts the sample. The graphene/hBN heterostructure is transferred onto this pyramid using a polymer transfer method that never directly touches the graphene (see Methods for details). 

An applied voltage reproducibly rotates the central rod by up to $\pm$10\(^\circ\), and the rod can additionally move in the Z direction by roughly $2$ \SI{}{\micro\meter} for a controlled approach to the sample (Fig. \ref{fig:1}e). Because the $\pm$10\(^\circ\) rotation range alone cannot reach both the $30^\circ$-aligned and $0^\circ$-aligned (Bernal) tBG configurations, we additionally separated the MEMS from the substrate and manually rotated it between measurements, defining three configurations (Configurations 1--3) that together span the full angle range. Relative to Configuration 1, Configuration 2 is rotated by $3\degree$, and Configuration 3 by $30 \degree$ (Fig. \ref{fig:1}f).

The heterostructures on the MEMS and on the substrate are fabricated separately and then combined using thermally cured glue. The assembled device is then wire-bonded to provide electrical connections for controlling the MEMS (Fig. \ref{fig:1}b). Optical interference between the graphene-covered center rod and the substrate helps determine the distance between the pyramid and the substrate \cite{SIline}. The pyramid was brought into contact with the substrate slowly, advancing only a few nanometers per step until it visually snapped onto the substrate (Fig. \ref{fig:1}c-d). This procedure allowed the contact pressure to be carefully controlled without applying excessive compressive force to the interface. Raman spectra were measured during the approach, as shown in Fig. \ref{fig:1}g-h (see also \cite{SIline}). 

To convert the relative rotation set by the MEMS into an absolute twist angle between the graphene layers, we identified two mirror-symmetric reference orientations using Raman spectroscopy. For a hexagonal lattice material like graphene, both $0\degree$ and $30\degree$ are mirror-symmetric alignments and therefore serve as fixed calibration points. Combining these reference points with the high-precision in situ rotation measurement and the manual rotation angle determined by optical microscopy allows us to determine the absolute twist angle at each measurement point in the study (Fig. \ref{fig:E_symmetrical}).

The study is organized as follows. First, we confirm the formation of a moir\'{e} potential at the twisted interface. We also test the robustness of the interface under compressive and decompressive forces. The moir\'{e} potential activates the graphene longitudinal optical (LO) phonon mode in Raman scattering \cite{CarozoNanolett,tBGRamanreview,venezuela}, allowing a precise mapping of the LO phonon mode that follows the momentum of the moir\'{e} potential. We then focus on the double-resonant 2D mode, which is sensitive to the electronic band, to study the band hybridization as a function of twist angle. Finally, the phonon band hybridization near $0^\circ$ is examined through the $\Gamma$-point phonon mode known as the G mode \cite{NFPmodel, NFP_WSe2}.

\subsection{Formation of the moir\'{e} potential at the engaged interface}

Bringing two independently fabricated graphene layers into contact can introduce strain and imperfect adhesion, whether from nanofabrication residue \cite{AFMcleangr,mote2afm,granneal,SIline}, airborne contaminants \cite{airborneWCA, AdvMatWCA} or humidity \cite{watercontrolgap}. This risk is amplified by the relatively large contact area ($\sim3.5\times3.5\text{ $\mu$m}^2$), which increases the likelihood of wrinkles or bubbles forming at the interface. Confirming that the interface remains robust across repeated contacts is therefore an essential prerequisite for this study.

\begin{figure*}[ht]
\includegraphics[width=6.5 in]{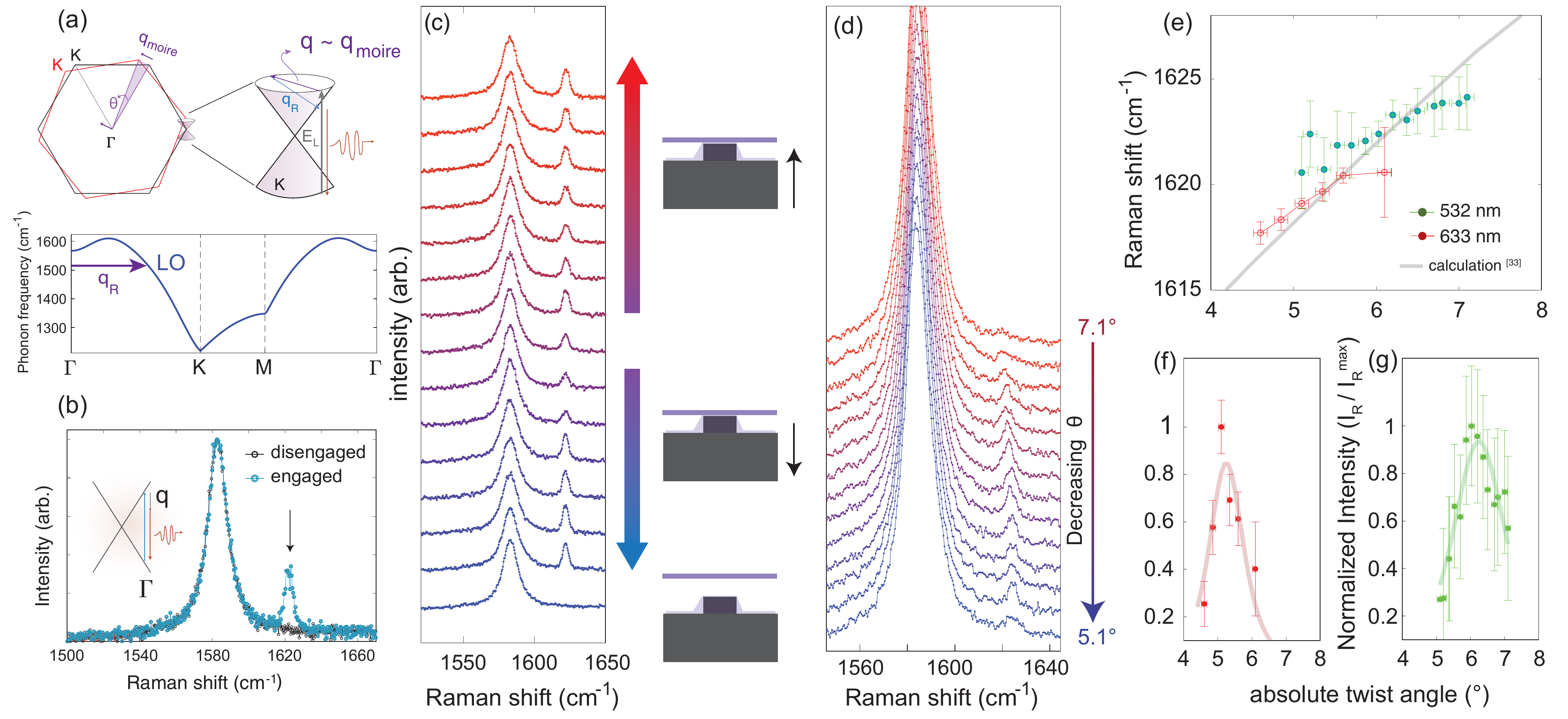} 
\centering
\caption{\textbf{Direct measurement of the moir\'{e} potential and graphene LO phonon}
\textbf{(a)} Raman scattering process for the ``R'' mode and calculated LO phonon dispersion from density functional theory.
\textbf{(b)} Raman G-mode spectra measured with (blue) and without (black) engagement at a relative rotation angle of approximately $6.6 \degree$ between the two graphene lattices. The black arrow indicates the emergence of a moir\'{e}-induced peak (R mode) arising from an additional Raman process where the electron scatters off the superlattice potential when the MEMS is engaged with the substrate. The inset shows schematic of the Raman mechanism for the graphene G  mode \cite{ferrari_review}. 
\textbf{(c)} G-mode and R-mode spectra measured while pushing the pyramid into the substrate with excess voltage and while retracting after engagement. Due to the vdW force between the two layers, the graphene layers remain in contact during retraction. The R peak is robust throughout the MEMS Z-direction operation. The bottom-most spectrum corresponds to the point at which the two graphene layers are fully separated.
\textbf{(d)} Angle-dependent evolution of the R peak. The twist angle between the two graphene layers varies from $7.1\degree$ (red) to $5.1\degree$ (blue) in steps of $0.1^\circ$ - $0.17^\circ$. 
\textbf{(e)} Fitted center Raman frequency of the R mode measured with $532$ nm (green) and $633$ nm (red) laser excitation. The grey line represents the dispersion predicted from calculations of the single-layer graphene LO phonon dispersion \cite{venezuela}. Due to systematic uncertainty in the overall scale of the DFT calculations, the grey curve from \cite{venezuela} has been shifted by 20 cm\(^{-1}\) \cite{CarozoNanolett}. 
\textbf{(f,g)} Intensity of the R mode for measurements using (f) $633$ nm and (g) $532$ nm excitation. The x-axis represents the absolute twist angle between the two graphene lattices, determined from the measurements in Fig.\ref{fig:E_symmetrical}. The smooth lines (red for $633$ nm and green for $532$ nm) are fits \cite{dresselhausGband}. The fitted Fermi velocities of the graphene are $1.2\pm0.3\times10^6 \text{ m/s}$ and  $1.2\pm0.9\times10^6 \text{ m/s}$ for the $633$ nm and $532$ nm measurements, respectively. 
}
\label{fig:2}
\end{figure*}

We demonstrate that a robust moir\'{e} interface forms when the two graphene surfaces are brought into contact (Fig. \ref{fig:2}b). The black and blue traces show the G-mode Raman spectra before and after the MEGA2D pyramid is brought into contact with the substrate, respectively. A prominent peak near $1620 \text{ cm}^{-1}$ appears upon engagement at a rotation angle of $6.6^\circ$ ($532$ nm excitation) and is robust under multiple engagements and different compressive strains (Fig. \ref{fig:2}c). The peak persists under decompressive pressure, with no significant shift in its position, because the vdW force continues to hold the two surfaces together. The peak disappears only once the two surfaces are fully separated.\\
This additional peak near a twist angle of $6^\circ$ (for $532 \text{ nm}$ excitation) is referred to as the ``R" peak \cite{CarozoNanolett}. The R peak is activated by the additional momentum transfer induced by the superlattice potential. When the momentum supplied by the superlattice potential ($q_{moir\acute{e}}$) matches the momentum of the graphene LO phonon, the R mode is activated (Fig. \ref{fig:2}a). Because this geometrical constraint directly links  the phonon momentum to the superlattice wavevector, the R-mode frequency disperses with twist angle. The momentum of the R mode can be expressed as a function of the twist angle as follows:
\begin{equation}
    q_{moir\acute{e}} = \frac{8\pi}{\sqrt{3}a} \sin{\frac{\theta}{2}} \simeq q_R
\label{eq:qtheta}
\end{equation}
Here, $q_R$ denotes the momentum of the LO phonon, $a$ is the graphene lattice constant and $\theta$ is the twist angle between the top and bottom graphene layers. This is a resonant process, so the R mode is visible only when laser excitation promotes an electron to the conduction band and the condition in Eq. \ref{eq:qtheta} is satisfied. Assuming a linear band dispersion of the Dirac cone, which is a good approximation at large twist angles, the resonant condition is as follows: 
\begin{equation}
    E_L = 2\hbar \nu_F q(\theta)  = \hbar \nu_F  \frac{8\pi}{\sqrt{3}a} \sin{\frac{\theta}{2}} \label{eq:eLeq}
\end{equation}
Here, $E_L$ is the laser excitation energy and $\nu_F$ is the Fermi velocity. To extend the measured angular range, we performed additional measurements using $633$ nm laser excitation (Fig. \ref{figE:Rpeak633}). The fitted center frequencies of the R mode as a function of the twist angle are presented in Fig. \ref{fig:2}e.

The fitted R-mode intensities are presented in Fig. \ref{fig:2}f,g and were fitted using the resonant condition:
\begin{equation}
    I_{R} = \left|\frac{M}{(E_L-E_\text{intra}-i\gamma)(E_L-E_\text{intra}-\hbar \omega_R -i\gamma)}\right|^2 \label{eq:Rintensity}
\end{equation}
Here, $\hbar\omega_R$ is the energy of the R-mode phonon, $E_{\text{intra}}$ is the energy difference between the initial and the excited states of the electron in this Raman process, and $M$ is the product of the matrix elements for the electron-photon and electron-phonon interactions (see \ref{SI:Rintensityfit} of \cite{SIline}, where we follow the fitting procedure of \cite{dresselhausGband}). The maximum intensities occur at twist angles of $5.25^\circ$ and $6.25^\circ$ for $633 \text{ nm} $ and $532 \text{ nm}$ excitation, respectively. The extracted Fermi velocity of single-layer graphene, $\nu_F$, is $1.2(3)\times10^6 \text{ m/s}$ for both $633 \text{ nm and } 532 \text{ nm}$ excitation. 

A precise mapping of the graphene LO phonon dispersion has not previously been achieved with a sufficient resolution and angular precision to enable detailed comparison with theoretical predictions \cite{venezuela,dresselhausGband,Erhardt_LOphonon}. 

Consequently, the dispersion obtained in this study (Fig. \ref{fig:2}e) provides a high-precision measurement of the graphene LO phonon dispersion. For comparison, the theoretical curve in Fig. \ref{fig:2}e is based on an ab initio phonon calculation from \cite{venezuela}, which has also been used in earlier works \cite{CarozoR, CarozoNanolett}.

\subsection{Electronic band hybridization in the tBG system} 

\begin{figure*}[ht]
\includegraphics[width=6.5in]{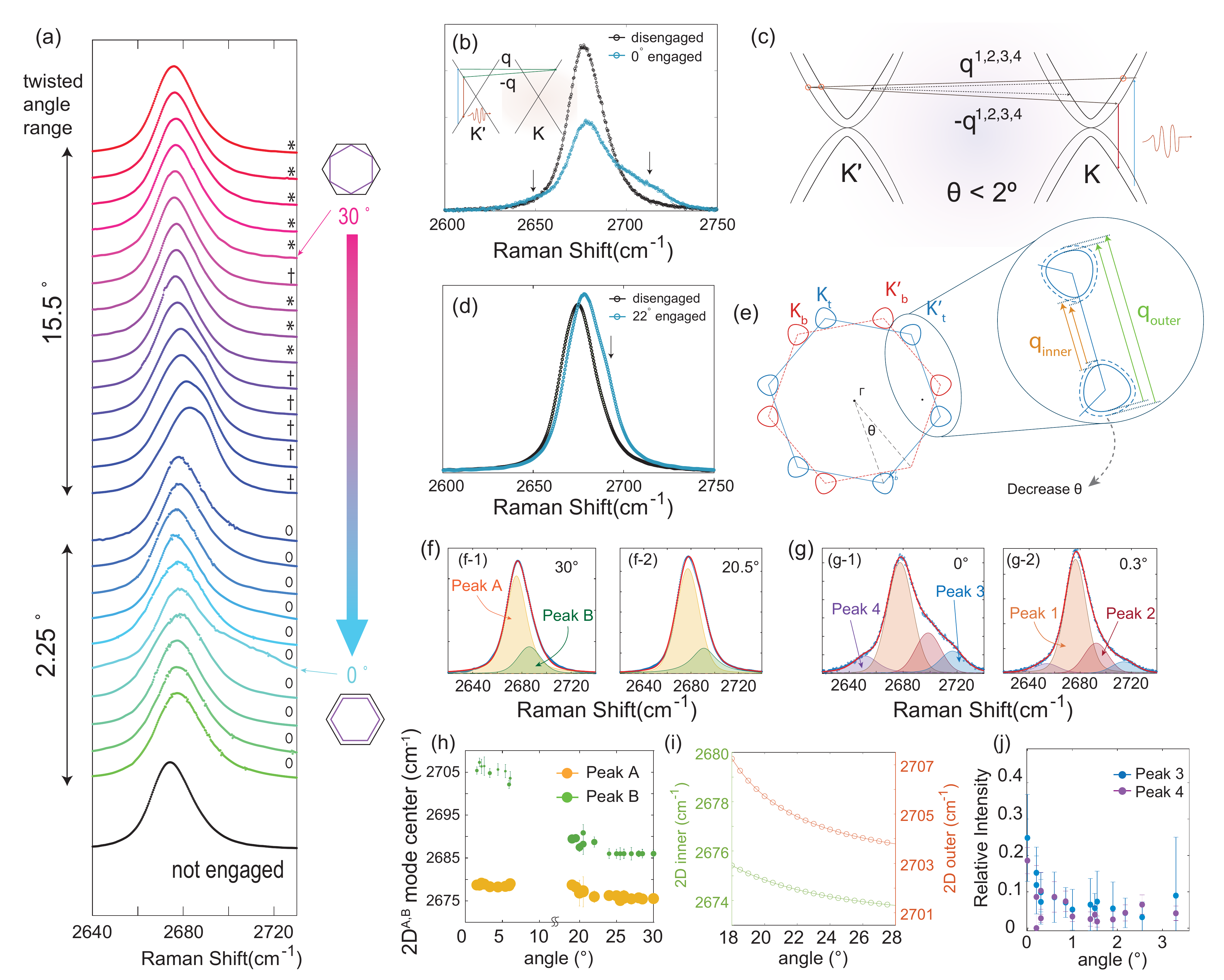} 
\centering
\caption{\textbf{Band hybridization of in situ rotated tBG}
\textbf{(a)} 2D-mode spectra obtained during in situ twisting for all three configurations. Spectra marked with an asterisk (*), a dagger (\(\dagger\)), and circles are from Configurations 1, 2, and 3, respectively. The twist angle range is indicated on the left. The rotation angle changes monotonically from the top to the bottom spectrum. The black spectrum shows the 2D mode when the MEMS is not engaged with the substrate. All spectra are normalized to the highest intensity of each spectrum for easier comparison.
\textbf{(b)} Raman 2D-mode spectra measured with (blue) and without (black) engagement at $0 \degree$ relative rotation between the two graphene lattices. The black arrows indicate new sub-peaks that appear due to band hybridization near the K point. The maximum 2D-mode intensity decreases by $44\%$ due to emergent resonant modes. The inset shows schematics of the Raman mechanisms for the graphene 2D mode \cite{ferrari_review}. 
\textbf{(c)} Schematics for Raman scattering associated with the 2D mode in a graphene bilayer with a small twist angle (\(\theta<2^\circ\)) \cite{zettlraman}. Due to band hybridization between two Dirac cones very close to each other, four different phonon momenta, $q_{1,2,3,4}$, can couple and activate four different double-resonant conditions.
\textbf{(d)} Raman 2D-mode spectra measured with (blue) and without (black) engagement at $22 \degree$ relative rotation between the two graphene lattices.
\textbf{(e)} Schematics of the inner-outer process of the 2D mode \cite{Inner_outer_tonyheinz,venezuela,Inner_outer_scirep}. Two layers are overlaid to illustrate the twist angle, $\theta$. The inner (orange arrow) and outer process (green arrow) are both present in the 2D mode, broadening the Lorentzian peak and producing asymmetry. As the twist angle changes, the electronic bands hybridize near the K point and affect the two processes by modifying the scattering momentum.
\textbf{(f)} 2D Raman spectra and fitted curves at twist angles of \(30^\circ\) and \(20.5^\circ\). The two sub-peaks, Peaks A and B, arise from the inner-outer processes of the 2D mode \cite{Inner_outer_tonyheinz,Inner_outer_scirep,venezuela}.
\textbf{(g)} 2D Raman spectra at a twist angle of \(0^\circ\) and \(0.3^\circ\) from the third configuration. The small-angle 2D modes are fitted with a four-sub-peak model based on the mechanism expressed in (c). 
\textbf{(h)} The fitted peak center for Peaks A and B for twist angle larger than \(1^\circ\).  
\textbf{(i)} Raman shift calculated from a continuum model with trigonal warping \cite{BMmodel}.
\textbf{(j)} The relative intensity of Peaks 3 and 4. The intensity increases as the angle approaches \(0^\circ\).
}
\label{fig:3}
\end{figure*}

The Raman 2D-mode spectra are presented over a wide range of twist angles spanning the regions near $30^\circ$ and $0^\circ$ (Fig. \ref{fig:3}a). Near $30^\circ$, the maximum intensity of the 2D mode is not expected to change substantially because the electronic states involved in the double-resonant Raman process are only weakly affected by the interlayer band hybridization. In contrast, the line shape of the 2D mode becomes increasingly sensitive to band hybridization as the twist angle decreases. As shown in Fig. \ref{fig:3}h and Fig. \ref{fig:E_rotation}, the 2D mode exhibits an overall blueshift accompanied by an increasing asymmetry. The asymmetry originates from the inner and outer double-resonant Raman processes, analogous to the asymmetric 2D line shape observed in monolayer graphene due to trigonal warping \cite{venezuela,trigonalwarping_TB,Inner_outer_scirep,Inner_outer_tonyheinz}. The individual Raman processes are illustrated in Fig. \ref{fig:3}e. The two components arising from inner and outer scattering processes, Peak A and Peak B (Fig. \ref{fig:3}f,h), both blueshift as the twist angle decreases, reflecting the modification of the electronic dispersion near the K point due to interlayer band hybridization. As the twist angle decreases, the equal-energy band contour corresponding to \(\pm E_L/2\) changes shape, shifting the inner and outer scattering phonon momenta farther from the \(K\) point. This effect is analogous to changing the laser excitation energy in monolayer graphene, which probes different equal-energy contours \cite{2Ddispersion,venezuela,Inner_outer_tonyheinz}. 

The peak positions change rapidly near \(20^\circ\), where the electronic bands near \(\pm1.2\text{ eV}\) exhibit strong hybridization (Fig. \ref{fig:E_bandstructure}). The expected Raman shift as a function of rotation angle, calculated as described in \cite{SIline}, is presented in Fig. \ref{fig:3}i. The two-peak model remains sufficient over the range \(3^\circ<\theta<10^\circ\), because the relevant resonant electronic states are modified primarily at energies below the laser excitation energy, \(E_{L} = 2.41 \text{ eV}\). Importantly, using a single device decouples the evolution of the asymmetric spectral line shape from extrinsic perturbations such as strain and defects, enabling a systematic investigation of the intrinsic modification of the electronic band structure with twist angle \cite{defect_2D,2dbandsplit}.

Near \(0^\circ\), the hybridized electronic bands formed near the K point enable multiple resonant pathways for the 2D-mode Raman process (Fig. \ref{fig:3}c). The resulting multiple Raman channels reduce the maximum 2D-mode intensity by \(\sim\)44\(\%\), consistent with previous studies of bilayer graphene \cite{zettlraman,ferrari_1,ferrari_review}. The Raman spectra with the four-peak fitting are presented in Fig. \ref{fig:3}g. The peak near $2640 \text{ cm}^{-1}$ (Peak 4), the characteristic 2D-mode response of bilayer graphene and its hybridized electronic band structure \cite{barbosa,anomaliesRaman}, becomes particularly pronounced near the Bernal configuration (Fig. \ref{fig:3}j). Both additional peaks, Peak 3 and Peak 4 (Fig. \ref{fig:3}g), decrease in intensity and become indistinguishable from the main peak as the twist angle increases toward \(2^\circ\), demonstrating the strong sensitivity of the 2D-mode line shape to the evolution of the hybridized electronic band structure. A closer examination of the Raman spectrum in the Bernal configuration (Fig. \ref{fig:3}g-1) reveals a 2D line shape that differs from those reported for pristine bilayer graphene \cite{ferrari_1,barbosa}. This difference arises from the finite pyramid area (\(\sim 3.5 \times 3.5\) \SI{}{\micro\meter^2}), which places a monolayer-bilayer boundary close to the pyramid center where the Raman spectra were measured (see \ref{SI:grapheneboundaryraman} of \cite{SIline} and Refs. \cite{edgeferrari,edgeRaman}).

\subsection{Effect of the moir\'{e} potential on G-mode phonon hybridization}

\begin{figure*}[ht]
\includegraphics[width=6in]{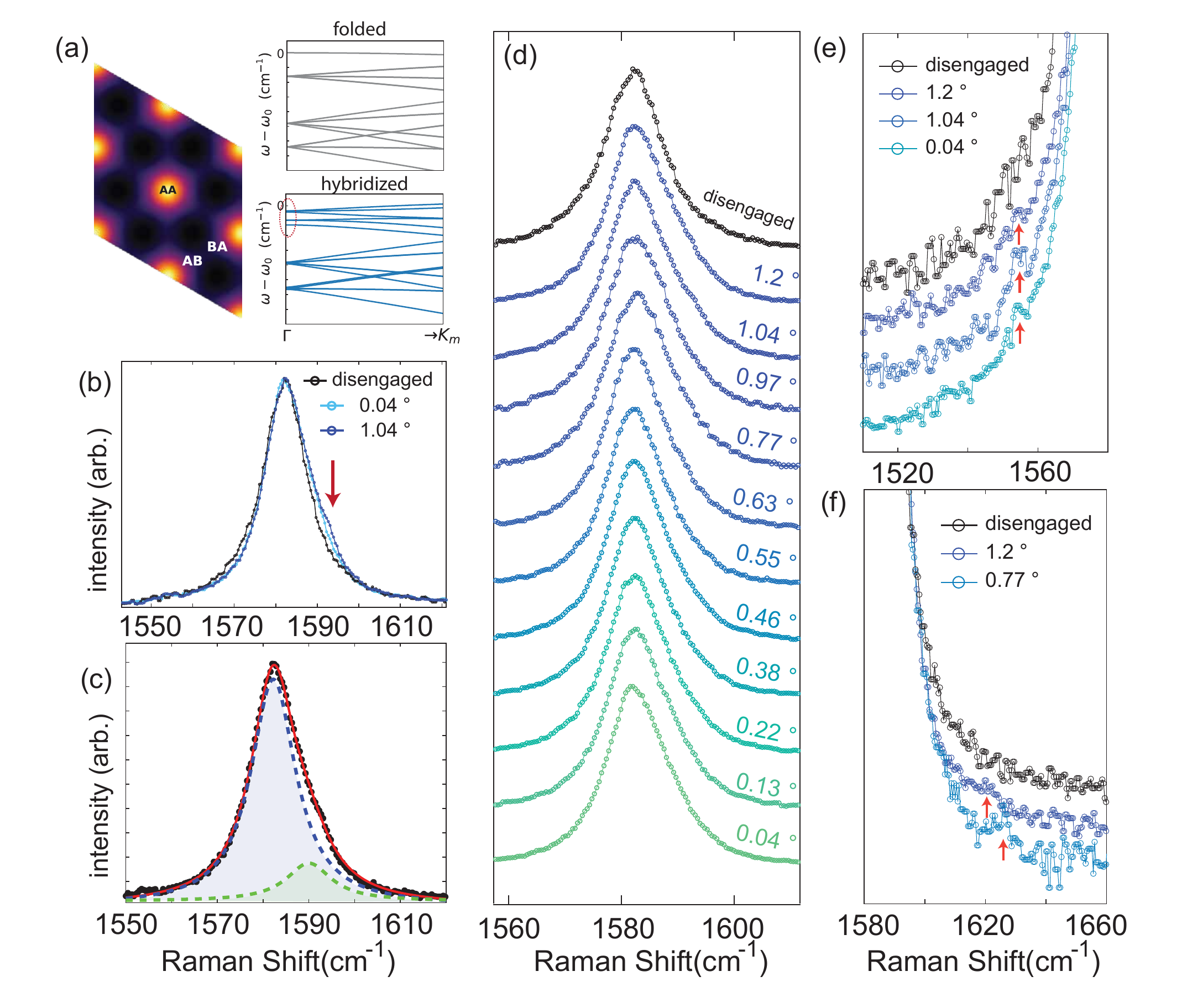} 
\centering
\caption{\textbf{Phonon band hybridization via the moir\'{e} superlattice}
\textbf{(a)} Schematics for phonon-band folding and hybridization induced by the moir\'{e} superlattice \cite{NFPmodel}. 
\textbf{(b)} Selected G-band spectra at twist angles \(0.04^\circ\) and \(1.04^\circ\), and with the MEMS disengaged. The asymmetric line shape develops as the MEMS is engaged with the substrate and the hump around \(1590 \text{ cm}^{-1}\) is more prominent around \(1.04^\circ\).
\textbf{(c)} Fit of the \(1590 \text{ cm}^{-1}\) side peak for \(1.04^\circ\).
\textbf{(d)} Full G-band spectra near \(0^\circ\). The light blue spectrum corresponds to the MEMS disengaged from the substrate. The additional peak near \(1590\text{ cm}^{-1}\) is most developed near \(1.04^\circ\). The feature does not fully disappear until \(\sim 2^\circ\) (see \cite{SIline}).
\textbf{(e)} Zoomed-in G-band spectra near \(0^\circ\) to show small peak features. The red arrows indicate the moir\'{e}-modulated side peaks. The center shifts slightly with rotation angle. Small peaks around \(1560 \text{ cm}^{-1}\) are present over the full \(1.2^\circ\) range, while the peak around \(1620\text{ cm}^{-1}\) is most prominent near \(0.77^\circ\). All spectra are normalized to the maximum intensity.
\textbf{(f)} Zoomed-in Raman spectra near the G band. The red arrows indicate the moir\'{e}-modulated side peaks around \(1560\text{ cm}^{-1}\) and \(1620\text{ cm}^{-1}\). The \(1620\text{ cm}^{-1}\) side peak is most prominent around \(0.77^\circ\).
}
\label{fig:4}
\end{figure*}

Unlike the 2D mode, which is sensitive to electronic band hybridization near the K point, the G mode exhibits signatures of phonon band hybridization involving \(E_{2g}\) phonons near the $\Gamma$ point. First, for small twist angles (less than \(\sim2^\circ\)), lattice relaxation produces well-defined regions of local AA, AB/BA, and saddle-point (SP) stacking, as shown in Fig. \ref{fig:4}a \cite{latticerelaxation}. Due to different interlayer interactions, each domain can host distinct local phonon modes \cite{youngwoosonPRB, nanoraman,nanoraman2,NFPmodel}. Second, the moir\'{e} potential introduces an additional periodicity that folds the graphene phonons into the smaller moir\'{e} Brillouin zone. The folded finite-momentum phonon branches can hybridize at the \(\Gamma\) point, producing sidebands near the original G mode. Figure \ref{fig:4}a uses a toy model with a single parabolic branch to illustrate phonon folding and hybridization induced by the moir\'{e} potential \cite{NFPmodel,SIline}. The circles on the branches illustrate the phonon modes that hybridize near the G mode and produce sub-peaks. Both effects originate from the modulation of the interlayer coupling by the moir\'{e} potential and both modify the Raman response near the G mode in tBG. 
 
Among the multiple sub-peaks that surround the G mode, the feature near \(1590 \text{ cm}^{-1}\) (Fig. \ref{fig:4}b-c)--previously observed in tBG samples and attributed to phonon band folding and hybridization \cite{anomaliesRaman,nanoraman}--contributes to the asymmetric G-mode line shape and is observed over a twist-angle range from \(0^\circ\) to \( \sim 3^\circ\). Smaller sub-peaks around \(1560 \text{ cm}^{-1}\) are also present in the same angular range, although they are more difficult to resolve because of the broad central G peak (Figs. \ref{fig:4}c and \ref{fig:E_Gbandrotation}). Nevertheless, these features can be captured by including additional weak peaks in the fitting of the spectra (Fig. \ref{fig:E_Gbandrotation}). In contrast, the feature near \(1620\text{ cm}^{-1}\) is particularly prominent at \(0.77^\circ\) (Fig. \ref{fig:4}f). A similar feature was observed near the SP regions in small-angle tBG \cite{nanoraman2}. 

\subsection{Discussion and Outlook}

We demonstrated a robust and reproducible tBG device in which the twist angle can be tuned in situ. We probed the properties of the tBG system by tracking the evolution of Raman modes under continuous angle rotation. In particular, we explored the effect of the superlattice potential on both the electronic band structure and the phonon dispersion, and we precisely mapped the graphene LO phonon dispersion by examining the moir\'{e}-induced R peak. By providing a single, angle-tunable, and reproducible platform, our approach reduces device-to-device variability and enables a precise mapping of graphene phonon spectrum and related graphene systems.

As a demonstration of light-matter interaction at an interface whose twist angle is controlled in situ, this work opens several directions for further study. Combining our angle-tunable device with tunable laser excitation could provide selective access to different phonon branches, such as transverse optical phonon modes \cite{miniarpestBG}. Moreover, the same platform could be adapted to cryogenic conditions, enabling further studies of electron-phonon interactions that could aid our understanding of the unconventional superconductivity in tBG \cite{yuanSC, elphGuinea, miniarpestBG,devereaux, ABernevig_phononSC}.

Finally, the reproducibility of the contact interface and the robustness of moir\'{e}-potential formation under varying compressive forces underscore the broader potential of mechanically controlled interface devices.
Our precise and deterministic twist-angle control enables, for the first time, direct in situ tuning of a reproducible and controllable hybridized band structure. These results establish the twist angle as a reliable tuning parameter and open a pathway toward systematic exploration of the full correlated phase diagram in twisted graphene and beyond \cite{zuipaper,KinFaiJieperspec}.

\bibliographystyle{naturemag}
\bibliography{references_ramanMEMS}

\section*{Data availability}
Source data are provided with this paper. All other data that support the findings of this study are available from the corresponding authors upon reasonable request.

\section*{Acknowledgments}
We thank Albert Zhu and Daniel Bennett for sharing data from their DFT calculations and Prof. Bertrand Halperin for helpful discussions. A.Y. acknowledges support from the Army Research Office under Grant number W911NF21-2-0147 and the Gordon and Betty Moore Foundation through Grant GBMF 12762. D.T.L and E.K. acknowledge support from the Simons Foundation Award No. 896626 and the U.S. Army Research Office (ARO) MURI project under Grant No. W911NF-21-0147. Y. C. acknowledges support by the US Department of Energy, Basic Energy Sciences, under award DE-SC0026044. Z. H. acknowledges support by Laboratory Directed Research and Development at the Lawrence Berkeley National Laboratory.

\section*{Author contributions}
S.P., A.Y. and Y.C. conceived and designed the experiments; Y.C. designed and fabricated the MEMS device; Y.C., S.P., and Z.H. developed graphene device fabrication process; S.P. fabricated graphene device on MEMS and graphene device on substrate; S.P. conducted the Raman measurement; S.P. and D.T.L. analyzed the result on Raman measurement; S.P. wrote the manuscript with input from A.Y., Y.C., D.T.L., Z. H., and E.K.

\section*{Competing interests}
The authors declare no competing interests.

\clearpage

\renewcommand{\thefigure}{E\arabic{figure}}
\renewcommand{\thesection}{E\arabic{section}}
\renewcommand{\thesubsection}{E\arabic{subsection}}
\renewcommand{\theequation}{E\arabic{equation}}
\renewcommand{\thetable}{E\arabic{table}}
\setcounter{figure}{0} 
\setcounter{equation}{0}
\onecolumngrid

\section*{Methods}
\subsection*{Graphene MEMS and substrate fabrication}
The MEMS device was fabricated following the protocol described in \cite{yuan2014onchip}. For both the MEMS-side and substrate-side graphene/hBN devices, a polypropylene carbonate (PPC) film on a PDMS stamp was used to first pick up hBN and then graphene. The PPC film was then carefully peeled from the PDMS stamp and manually flipped. Next, the film was placed onto a PDMS stamp containing a hole so that the flake would not come into contact with the PDMS after flipping. The film was subsequently released onto either the MEMS device or the substrate by heating the sample to $60\text{--}120\,^{\circ}\mathrm{C}$ while the film was in contact with the surface. After the transfer, the sample was annealed in vacuum at $300\text{--}350\,^{\circ}\mathrm{C}$ for 4–6 hours. Following annealing, both the substrate and MEMS devices were characterized and cleaned using AFM.

\subsection*{Placing MEMS on the substrate}
Micro silica beads were used to maintain at least 2 \SI{}{\micro\meter} distance between the MEMS pyramid and the substrate. The beads were bought from Applied Physics (Silica-Dry-2.5\(\mu\)m5g). The beads were mixed with a thermally curable glue (Crystalbond\(^{TM}\) 509) using cyclopentanone (VWR international INC.) and injected onto the surface of the MEMS with a nano-injector(Nanoject III, Drummond Scientific). The substrate was held on a PDMS stamp. The MEMS and substrate were aligned on a transfer stage and thermally equilibrated at 180 \(\degree\)C. After cooling to room temperature, the MEMS-substrate assembly was separated from PDMS.

\subsection*{Raman measurement}
All Raman measurements were performed using a Horiba multiline Raman spectrometer with an 1800 grooves/mm and a laser power below 5 mW. For more than 30 seconds of measurement time, the laser power was kept below 1 mW. The voltage signal used to control the MEMS rotational gear was generated by a digital-to-analog converter connected to the computer via MATLAB. The MEMS rotation is controlled using a three-phase voltage signal \cite{yuan2014onchip} with the maximum voltage of 90 V on the electrode.

\clearpage

\renewcommand{\thefigure}{S\arabic{figure}}
\renewcommand{\thesection}{Section \arabic{section}}
\renewcommand{\thesubsection}{\arabic{subsection}}
\renewcommand{\theequation}{S\arabic{equation}}
\renewcommand{\thetable}{S\arabic{table}}
\setcounter{figure}{0} 
\setcounter{equation}{0}
\setcounter{page}{1}
\onecolumngrid
\setcounter{secnumdepth}{1}
\section*{Supplementary Information}

\section{Raman measurement while approaching with MEMS devices}
\label{SI:ramanapproach}

\begin{figure}
\centering
\includegraphics[width = 5 in]{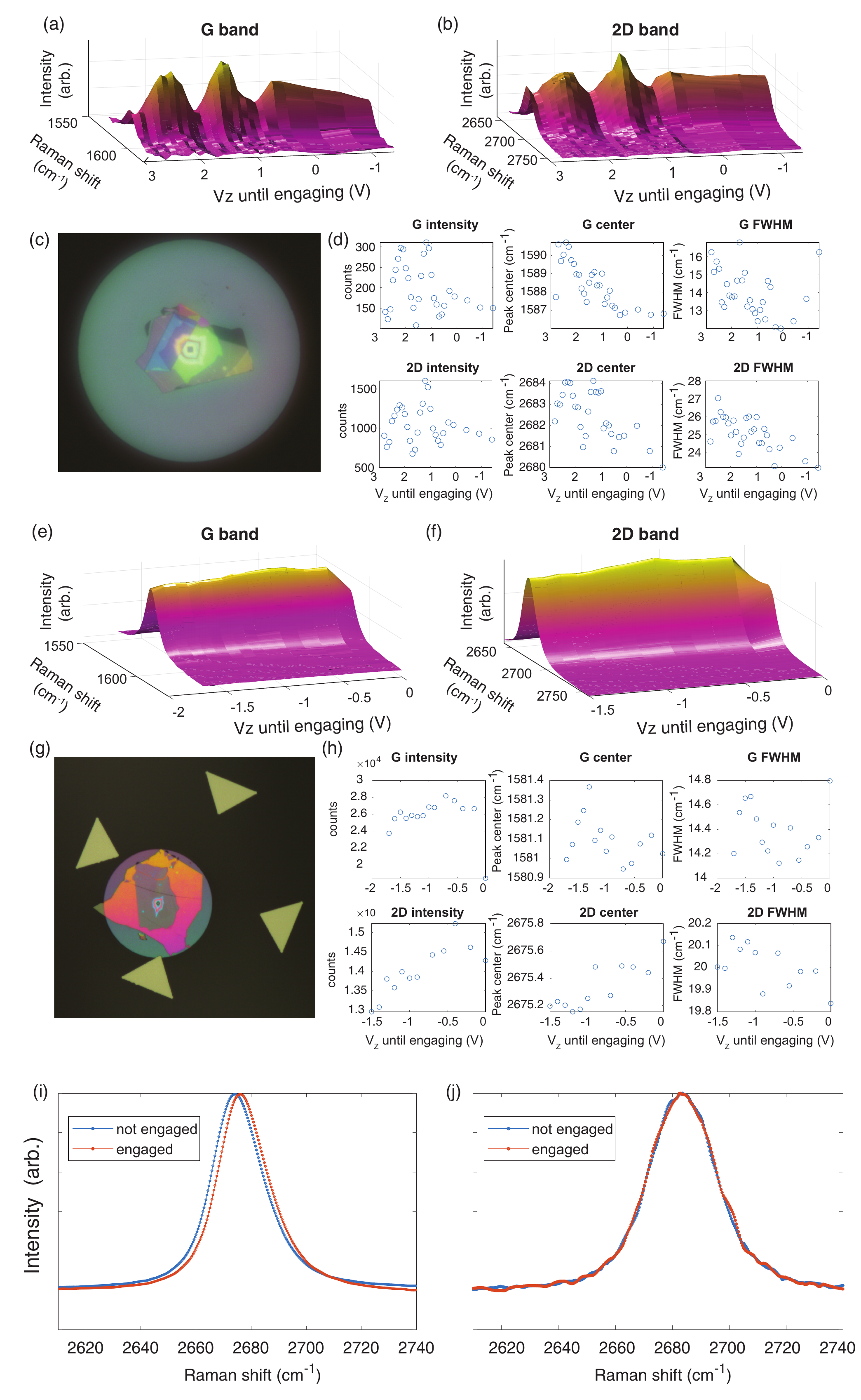}
\caption{\textbf{Raman G- and 2D- mode spectra for two different MEGA2D devices during approach to the substrate}
\textbf{(a,b)} G and 2D mode spectra as a function of Vz. Positive Vz corresponds to the MEMS being out of contact with the substrate, whereas negative Vz corresponds to the MEMS being in contact with the substrate and moving farther into contact.
\textbf{(c)} Optical image of device A. 
\textbf{(d)} Fitting result for G and 2D mode using Lorentzian function. The oscillation period for all three parameters is approximately 1 V in Z. After engagement (Vz = $0$ V), the oscillation is no longer observed. 
\textbf{(e,f)} G and 2D mode spectra for device B. 
\textbf{(g)} Optical image of device B.
\textbf{(h)} Fitting result for device B. The oscillation period remains approximately $1V$, but the oscillation amplitude is substantially smaller than that in (d).
\textbf{(i,j)} Comparison between just before engaging and after engaging for (i) device A and (j) device B. The intensity in each plot is normalized to easily compare the two spectra. The intensity of both spectra decreases as MEMS engages with the surface, but the lineshape of device A is identical before and after engaging. This is due to surface contamination of substrate side, possibly due to a thin layer of polymer.
}
\label{fig:E_approaching}
\end{figure}

In this section, we present Raman measurements acquired while approaching the MEMS (MEGA2D) pyramid to the substrate for two different devices. Device A (Fig. \ref{fig:E_approaching}a-d) and Device B (Fig. \ref{fig:E_approaching}e-h) are two separate substrate-MEGA2D assemblies, both incorporating hBN/graphene heterostructures. For Device A, the backside of the substrate was coated with a $10$-nm palladium-gold layer. For Device B, the backside of the substrate was not covered with $10$ nm of metallic layer. 

The oscillation of the Raman signal is very pronounced in Device A. The intensity and the center frequency oscillate as the MEMS approaches. The intensity oscillation can be understood as a consequence of optical interference between different interfaces. The oscillations in the G- and 2D-mode center frequencies and FWHM as a function of z can be understood as a consequence of cooperative Lamb shift and Purcell effect from the weak optical cavity formed by the reflective surfaces \cite{coorperativeLambshift_optica,gracelambshift, radiativelifetime}.
The oscillation period is approximately 1 V in Z voltage for both devices, however the oscillation amplitude of G and 2D mode are much larger in device A with metallic backside. Thus, the dominant reflection can be attributed to the metallic layer on the backside of the substrate, which strengthens overall `cavity'. The substantially larger cavity-induced effect in Device A suggests that this platform could be useful for studying coupling between excitons and the photon field \cite{radiativelifetime}.

Unlike the Device B, bringing the MEGA2D pyramid into contact with the substrate does not change the Raman spectrum of Device A (see Fig. \ref{fig:E_approaching}i,j). Further pushing the device does change the intensity of the Raman spectrum but not the lineshape itself. This indicates that interfacial hybridization did not occur in Device A upon contact, which is due to a thin layer of polymer that was left after annealing the sample. For Device B, AFM cleaning was performed after annealing. The lineshape of the 2D mode changes as the two surfaces are engaged (Fig. \ref{fig:E_approaching}).

\clearpage
\section{Angle reproducibility and sample cleaning}

\begin{figure}
\centering
\includegraphics[width = 5 in]{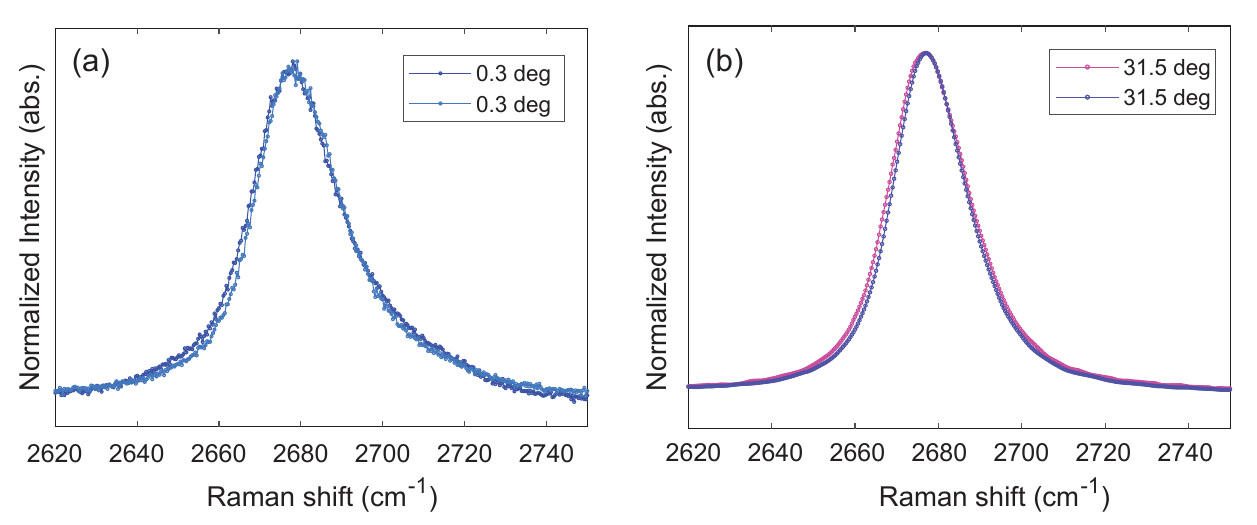}
\caption{
\textbf{Comparison of 2D-mode spectra between different experiment sets}
\textbf{(a)} 2D mode spectra of tBG at a twist angle of 0.3 \(^\circ\) from different days. The dataset is acquired from Configuration 3. 
\textbf{(b)} 2D mode spectra of tBG at a twist angle of 31.5 \(^\circ\) measured in Configuration 2. The two spectra were measured on different days.
}
\label{fig:E_reproducibility}
\end{figure}

Throughout the study, Raman spectra were acquired repeatedly at the same angle across different datasets to ensure the reproducibility of the data and the stability of the rotation. For example, Fig. \ref{fig:E_reproducibility} shows the reproduced data in Configuration 2 and 3. Even small twist angles below 0.5 \(^\circ\) are reproduced across multiple measurement sets. These measurements confirm precise angle control and the robustness of the tBG interface over repeated engagements.

\begin{figure}
\centering
\includegraphics[width = 7 in]{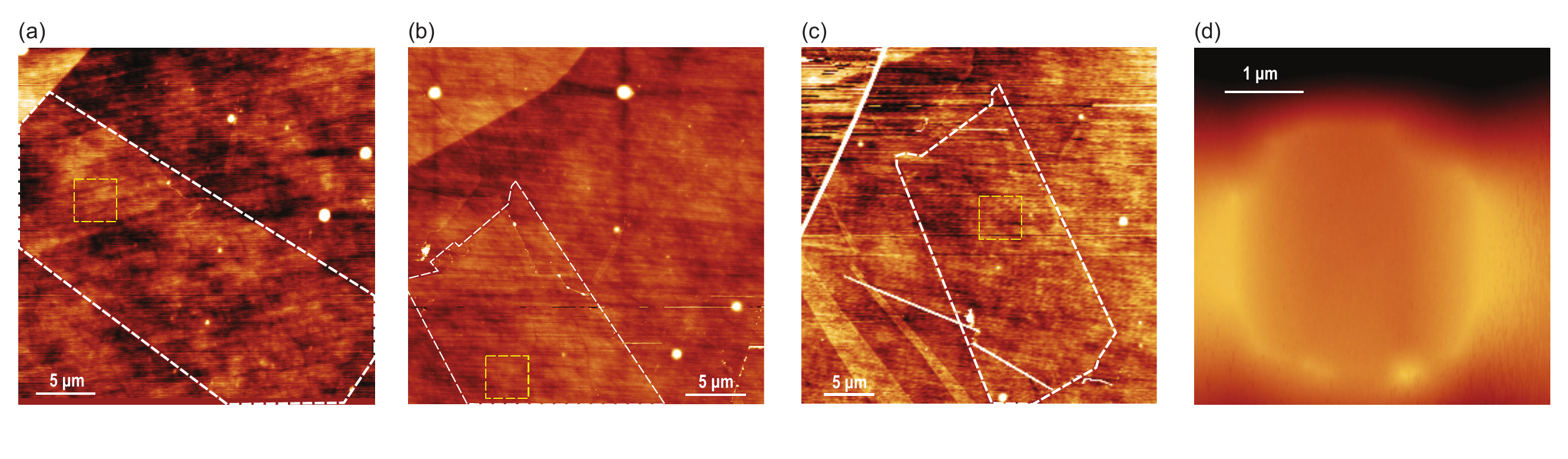}
\caption{
\textbf{AFM images of the substrate before Configuration 1-3 and of the MEMS graphene surface}
\textbf{(a-c)} AFM topography data before Configuration 1, 2, 3, respectively. The white line indicates the graphene layer on the substrate side, and the yellow square indicates where the MEMS pyramid engages to form the tBG interface. Scale bar is 5 \(\mu\)m.
\textbf{(d)} AFM topography data of the MEMS pyramid. The surface is extremely clean, with no visible debris or bubbles. The rounded-square-like feature in the middle is the extruded pyramid. The size of the pyramid is 3.5 \(\times\) 3.5 \(\mu\)m\(^2\). Scale bar is 1 \(\mu\)m.
}
\label{fig:E_AFM}
\end{figure}

As mentioned in the text, every time the substrate and the MEMS were separated, the substrate and the MEMS graphene were both inspected with AFM and cleaned to ensure a clean contact. The MEMS was separated from the substrate by heating the entire structure to 180\(^\circ C\), after which it was lifted with tweezers. Mechanical movement of the tweezers can cause the central rod to oscillate and contact the substrate. This unwanted contact could damage the substrate during the separation, but MEMS graphene tends to only collect debris from the substrate. The damage appears to occur only during separation of the MEMS from the substrate, as the Raman spectra remain reproducible (Fig. \ref{fig:E_reproducibility}). To ensure that each MEMS-substrate configuration forms a clean tBG interface, the MEMS was positioned over a clean area of graphene after AFM inspection. The AFM image obtained before each configuration and the corresponding tBG interface location are shown in Fig. \ref{fig:E_AFM}. The AFM image of the surface of the MEMS is also present in Fig. \ref{fig:E_AFM}. The size of the pyramid is 3.5 \(\times\) 3.5 \(\mu\)m\(^2\).

\begin{figure}
\centering
\includegraphics[width = 7 in]{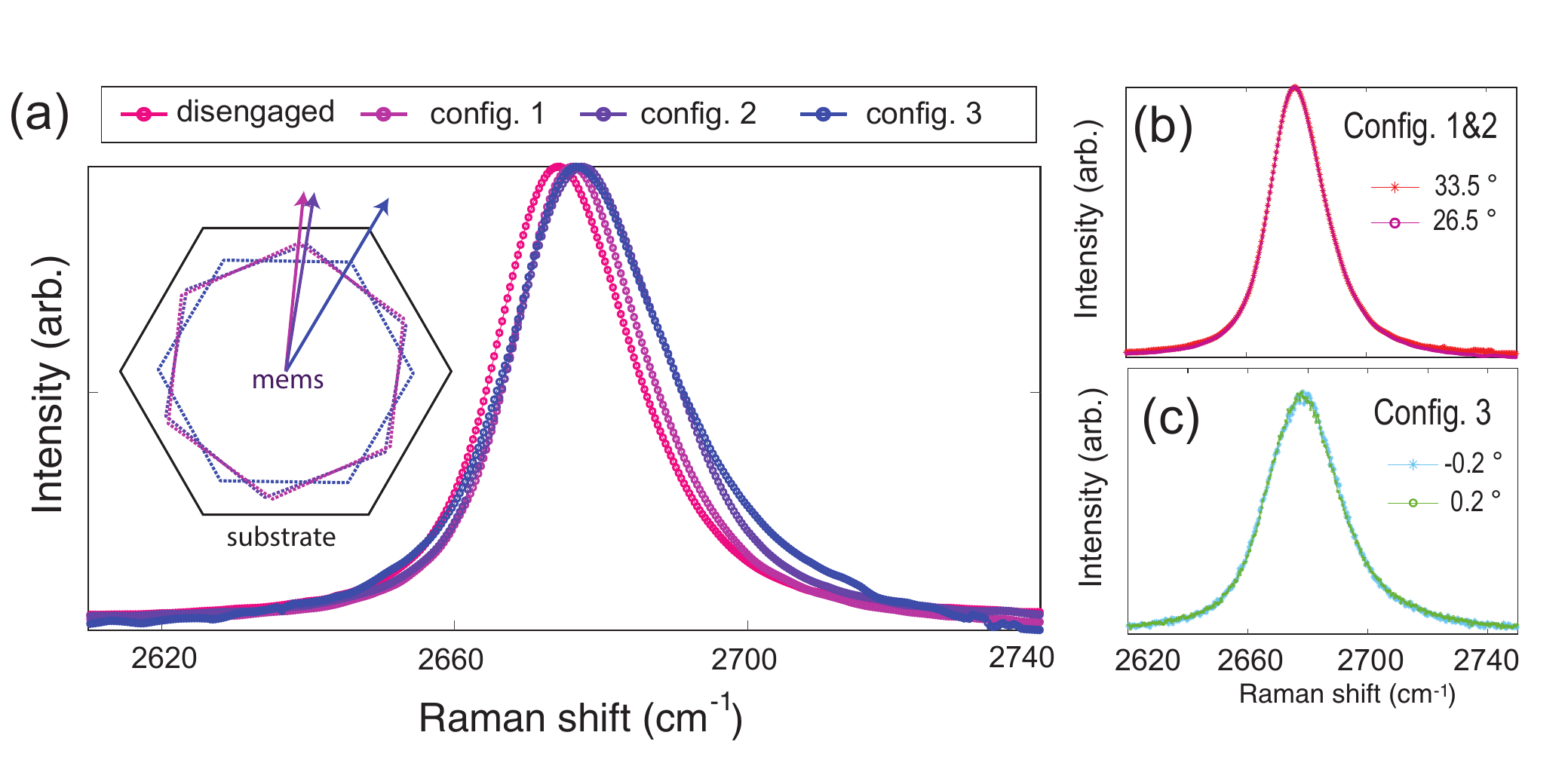}
\caption{
\textbf{2D mode for configuration 1-3 and raman spectra around symmetrical plane}
\textbf{(a)} Raman 2D-band spectra for all three configurations shown in (e), along with the disengaged state. The hexagons are oriented relative to the black hexagon (substrate) based on the in situ rotation and approximate manual rotation angle measurements. The first and second configurations are close to a $30\degree$ lattice rotation, whereas the third configuration is close to the Bernal bilayer configuration.
\textbf{(b,c)} Comparison of symmetric bands used to identify the high-symmetry points (\(30^\circ\) and \(0^\circ\)). The angles correspond to absolute twist angle between the two graphene layers.
}
\label{fig:E_symmetrical}
\end{figure}

\section{Full 2D Rotation Spectrum near \(0^\circ\)}
\label{SI:full_rotation}

\begin{figure}
\centering
\includegraphics[width = 7 in]{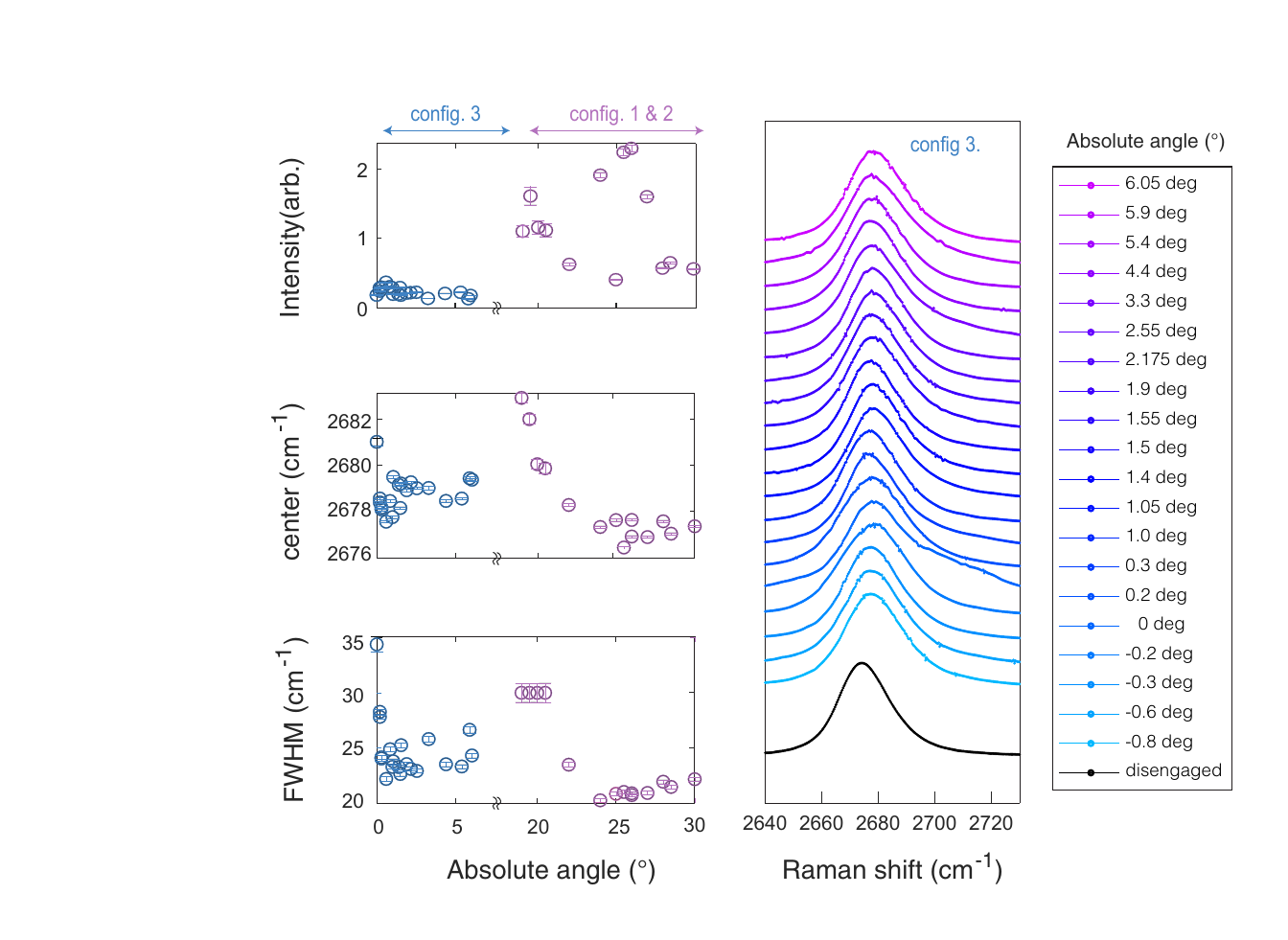}
\caption{\textbf{2D mode near $0\degree$ rotation and fitting Lorentzian for all rotation data}
\textbf{(a)} Fitted parameters for individual 2D spectra. Fits are performed using Lorentzian function \cite{zettlraman}.
\textbf{(b)} 2D mode spectra measured in configuration 3. The angles are calibrated based on the symmetry-related configurations as shown in Fig. \ref{fig:E_symmetrical}. 
}\label{fig:E_rotation}
\end{figure}

For angle calibration, two symmetry-related orientations were identified in Fig. \ref{fig:E_symmetrical}. 
Raman spectra were measured in configuration 3 for $0\degree$ to $6.65\degree$. Configuration 3 without any in situ twist is identified as $0.65^\circ$. The rotation near the first and second configurations, corresponding to tBG rotated by \(25^\circ\) and \(22^\circ\), respectively, is presented in the main text (Fig. \ref{fig:3}b). The intensity and FWHM, peak center were extracted from Fig. \ref{fig:3}b and Fig. \ref{fig:E_rotation} using Lorentzian fits for comparison with \cite{zettlraman}. The fitted values are symmetric around $0\degree$ and $30\degree$, confirming the accuracy of the calibrated angle. Following the fitted center frequency from 30\(^\circ\), the onset of the blueshift of 2D mode is around 20\(^\circ\). The FWHM of 2D mode also increases as the angle deviates more from 30\(^\circ\). These tendencies near 30\(^\circ\) are in accordance with the result presented in \cite{zettlraman}. The intensity near 0\(^\circ\), measured in Configuration 3, shows significant reduction compared to those near 30\(^\circ\). 

The side peak near $2710 \text{ cm}^{-1}$ does not reappear in this device, which has been observed in several studies \cite{barbosa,anomaliesRaman}. Nonetheless, the small peak near $2650 \text{ cm}^{-1}$ is still required to fit the 2D mode near $0\degree$, which is the signature peak for the band hybridization near K point for small-twist tBG \cite{anomaliesRaman}. 

\section{Peak fitting of  2D mode spectrum in the Bernal configuration}
\label{SI:peak_fit_2D}

\begin{figure}
\centering
\includegraphics[width = 7 in]{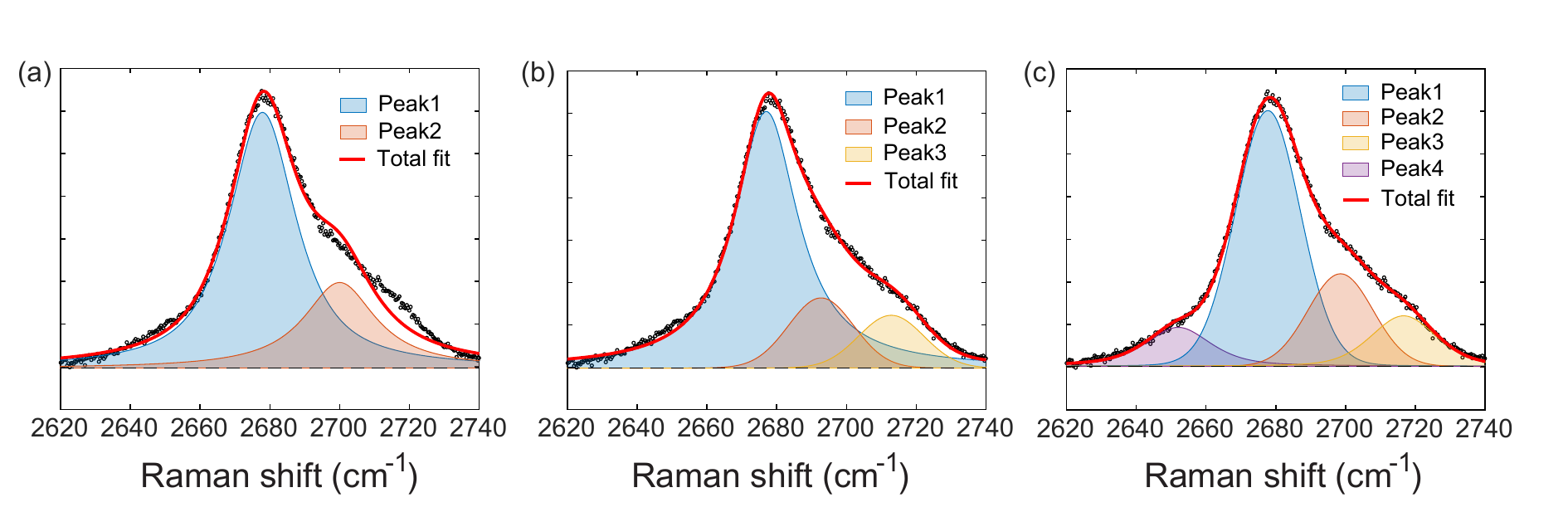}
\caption{\textbf{Sub-peak structure fitting of 2D mode at Bernal alignment}
\textbf{(a-c)} Fits using two-, three-, and four-peak models for the Raman spectrum of the Bernal structure. The sub-peaks are modeled using Voigt functions with independently fitted FWHM and peak intensities. The black dots represent the measurement data, and the dotted lines represent the fitted sub-peaks obtained using a curve-fitting tool from MATLAB. The total fitted line is the red line.
}
\label{fig:E_peakfitting}
\end{figure}

Here we present the sub-peak fitting of the Raman spectrum for the Bernal bilayer structure (Configuration 3, Fig. \ref{fig:3}e). We used a Voigt function for each sub-peak to minimize any unnecessary addition of peaks. A Lorentzian function is conventionally used to fit the spectrum for simplicity \cite{zettlraman}, but a Voigt function takes into account any interference between the scattering process and the environment and it is also used as an approximate form used in previous studies \cite{barbosa}. Because of the double-hump feature near 2710 $\text{cm}^{-1}$, more than three peaks are required to fit the spectrum. The small shoulder near 2640 $\text{cm}^{-1}$ is a signature for the band hybridization between two graphene layers at a small twist angle \cite{anomaliesRaman}. Fits using two-, three-, and four-peak models are presented (Fig. \ref{fig:E_peakfitting}). To capture the shoulders at \(2640 \text{ cm}^{-1}\) and \(2700\text{ cm}^{-1}\), a four-peak model is necessary.

\section{Rotation angle dependence of G band}
\label{SI:Gband_rotation}

\begin{figure}
\centering
\includegraphics[width = 7 in]{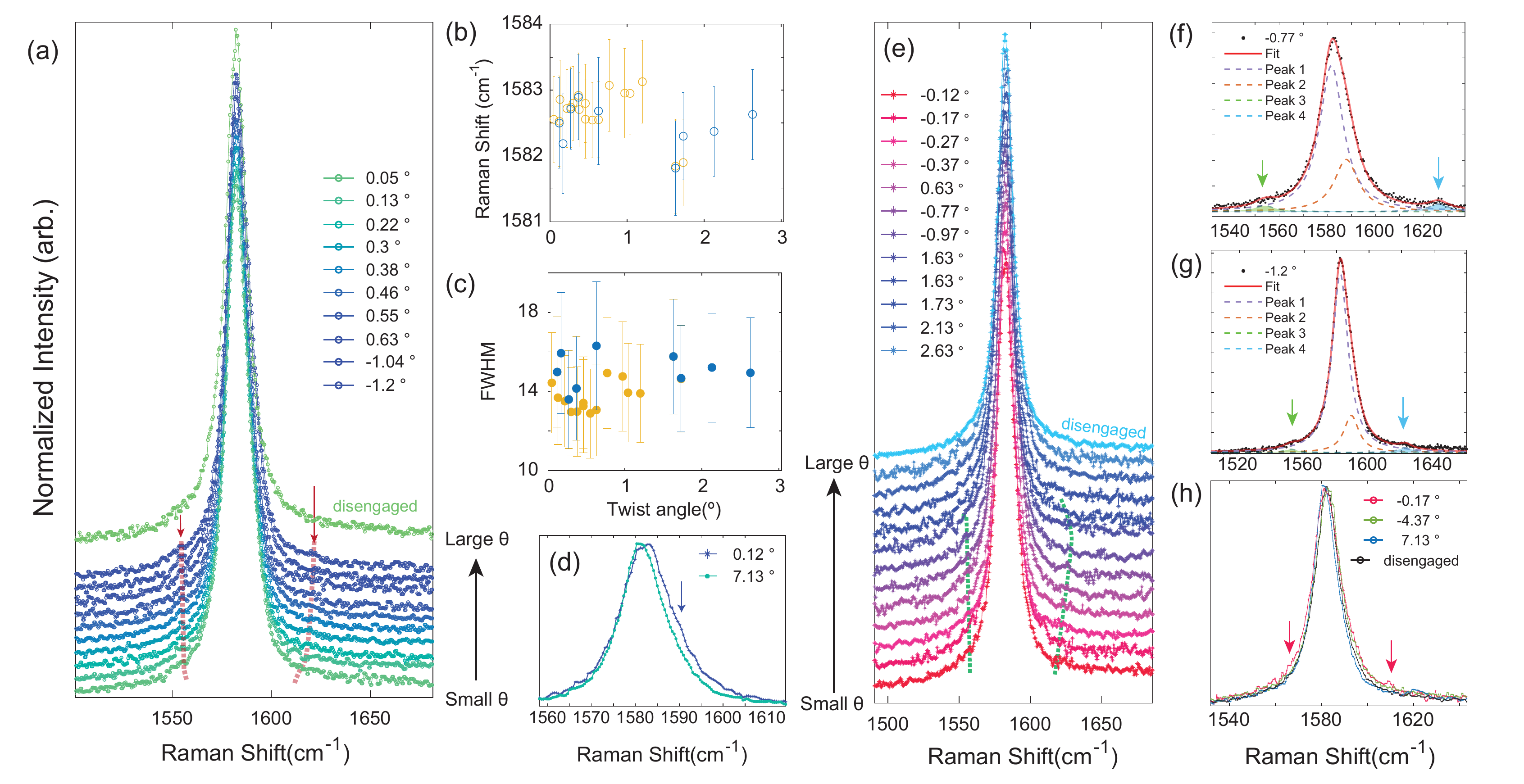}
\caption{\textbf{G band rotation-angle dependence}
\textbf{(a)} The same data from Fig. \ref{fig:4} with different spacing to emphasize the small peaks around $1550\text{ cm}^{-1}$ and $1610 \text{ cm}^{-1}$.
\textbf{(b,c)} Fitted G-mode center and FWHM for all the measurements. Slight increase of FWHM near $1^\circ$ is observed.
\textbf{(d)} G band measurement of \(0.12^\circ\) and \(7.13^\circ\) tBG. The arrow indicates the pronounced asymmetry in \(0.12^\circ\) spectrum.
The measurement is from the dataset shown in (e).
\textbf{(e)} Second measurement set acquired on different days over a larger rotation-angle range. Peaks around \(1520\text{ -- }1550\text{ cm}^{-1}\) are present up to \(2^\circ\). 
\textbf{(f,g)} Peak fitting for \(0.77^\circ\) and \(1.2^\circ\) spectra. Two peaks are used to fit the central G mode \cite{anomaliesRaman}. Small peaks (Peak 3 and 4) are for the small phonon side peaks fitted with Voigt function. 
\textbf{(h)} Zoomed out spectra for \(0.17^\circ\), \(4.37^\circ\), \(7.13^\circ\). There are small peaks around \(1560\text{ cm}^{-1}\) and \(1610 \text{ cm}^{-1}\) for small angles. The small peaks near the main peak are indicated as red arrows. The side peaks are not present for larger angles as shown for \(4.37^\circ\) and \(7.13^\circ\).
}
\label{fig:E_Gbandrotation}
\end{figure}

In addition to the data presented in Fig. \ref{fig:3}, G-band spectra at various rotation angles are shown in Fig. \ref{fig:E_Gbandrotation}. Two measurement sets presented in Fig. \ref{fig:E_Gbandrotation} were measured on different days. The peaks around the central G mode \(\sim 1583 \text{ cm}^{-1}\) that contribute to the asymmetry of the G mode disappear as the angle increases \(\theta\sim6^\circ\). The small peaks around \(1560 \text{ cm}^{-1}\) and \(1620\text{ cm}^{-1}\) measured in the main text persist until \(2.63^\circ\). Figure \ref{fig:E_Gbandrotation}h shows direct overlap of \(\theta=0.17^\circ\) Raman spectrum with those of \(\theta=4.37^\circ\) and \(\theta=7.13^\circ\) for comparison. The small peaks are not present at larger angles as can be seen in the figure. The fit for the small side peaks is also presented in Fig. \ref{fig:E_Gbandrotation}f and \ref{fig:E_Gbandrotation}g.

\section{nearly free phonon model}
The schematic in Fig. \ref{fig:4}a illustrates the nearly free phonon model, in which the moir\'{e} phonon spectrum is calculated from a single, simplified optical phonon branch. The G mode corresponds to the \(E_{2g}\) phonon modes, which is reduced to a single scalar optical branch in the model. For schematic purposes, we approximate the dispersion by a quadratic function centered around $\Gamma$. To emphasize the folding effect and the hybridization, we exaggerate the curvature of the dispersion such that the kinetic bandwidth over the moir\'{e} Brillouin zone is \(4~\text{cm}^{-1}\) as opposed to the actual bandwidth of  \(\sim10^{-2} \text{ cm}^{-1}\) for flat-band tBG. The effective Hamiltonian consists of the long-wavelength expansion of this optical branch and a periodic potential arising from the moir\'{e} superlattice. Expressed in a plane-wave basis of moir\'{e} reciprocal lattice vectors, analogous to the nearly free electron model, 
\begin{equation}
    H_{GG'}(k) = -\beta|k+G|^2\delta_{GG'}+V(G-G')
\end{equation}
where the potential \(V(G-G')\) is the Fourier transform of the local G-mode frequency shifts, which follows the stacking order as it varies from AA to AB/BA across the moir\'{e} cell. The eigenvalues of the Hamiltonian then yield the moir\'{e} phonon band structure, in which the folded branches hybridize through the moir\'{e} potential and open gaps at the mini-zone boundaries. The interlayer coupling does not appear explicitly in this single-band formulation, but only implicitly through the stacking dependence of the local phonon frequency that defines \(V(G-G')\). Lattice relaxation must also come into play as the stacking energy and the elastic energy compete with one another and locally deform the graphene lattice. This consideration is important for quantitative analysis; however, Fig. \ref{fig:4}a is intended only as a schematic illustration of the nearly free phonon model, a full calculation including relaxation is beyond the scope of this work. Claude was used to assist the development and refinement of code writing based on detailed instructions specifying detailed steps described in this section.

\section{Estimation of 2D mode twist angle dependence}
The 2D peak in monolayer graphene results from a double-resonance process in which electrons excited by the laser near the \(K\) valley are scattered to the \(K'\) valley and back again by a pair of opposite momentum iTO phonons. For a twisted bilayer with a relatively large twist angle \((\theta > 18^\circ)\), we assume that the 2D Raman spectrum arises primarily from individual graphene layers with the leading modification arising from changes in the electronic structure near the \(K\) point due to the presence of the other layer. To model the electronic structure for generic twist angles we use the Bistritzer-MacDonald (BM) model \cite{BMmodel} with a monolayer Hamiltonian augmented to include terms of quadratic order in momentum that are responsible for the trigonal warping \cite{trigonalwarping_TB}:
\begin{equation}
H_0(k) = h{\nu_F} [(k_x\sigma_x + k_y\sigma_y) - \frac{a}{4\sqrt{3}}((k_x^2 - k_y^2)\sigma_x - 2 k_x k_y \sigma_y)] 
\end{equation}
The interlayer hopping matrix $T$ is identical to that used in the original BM model, parametrized by hopping energy \(w = 110 \text{ meV}\). The resonance condition is satisfied at a given k-point when the energy difference between the conduction and valence bands matches the laser energy \((E_L = 2.33 \text{ eV})\). We search along a line in momentum space passing through adjacent \(K\) and \(K'\) valleys and identify the k-points in the electronic BZ that satisfy the resonance condition and calculate the corresponding momentum vectors for the dominant \(\Gamma-K\) (inner) and \(K-M\) (outer) phonon scattering vectors q, identified in \cite{venezuela}, that connect the \(K\) and \(K'\) valleys. Finally, using the monolayer phonon spectrum, the corresponding energy of the iTO phonon branch with momentum q yields the predicted 2D Raman shift as a function of twist angle shown in Fig. \ref{fig:3}i. Claude was used to assist with the development and refinement of code used in this work.

\section{Raman spectroscopy of graphene near bilayer-monolayer graphene boundary}
\label{SI:grapheneboundaryraman}
\begin{figure}
\centering
\includegraphics[width = 7 in]{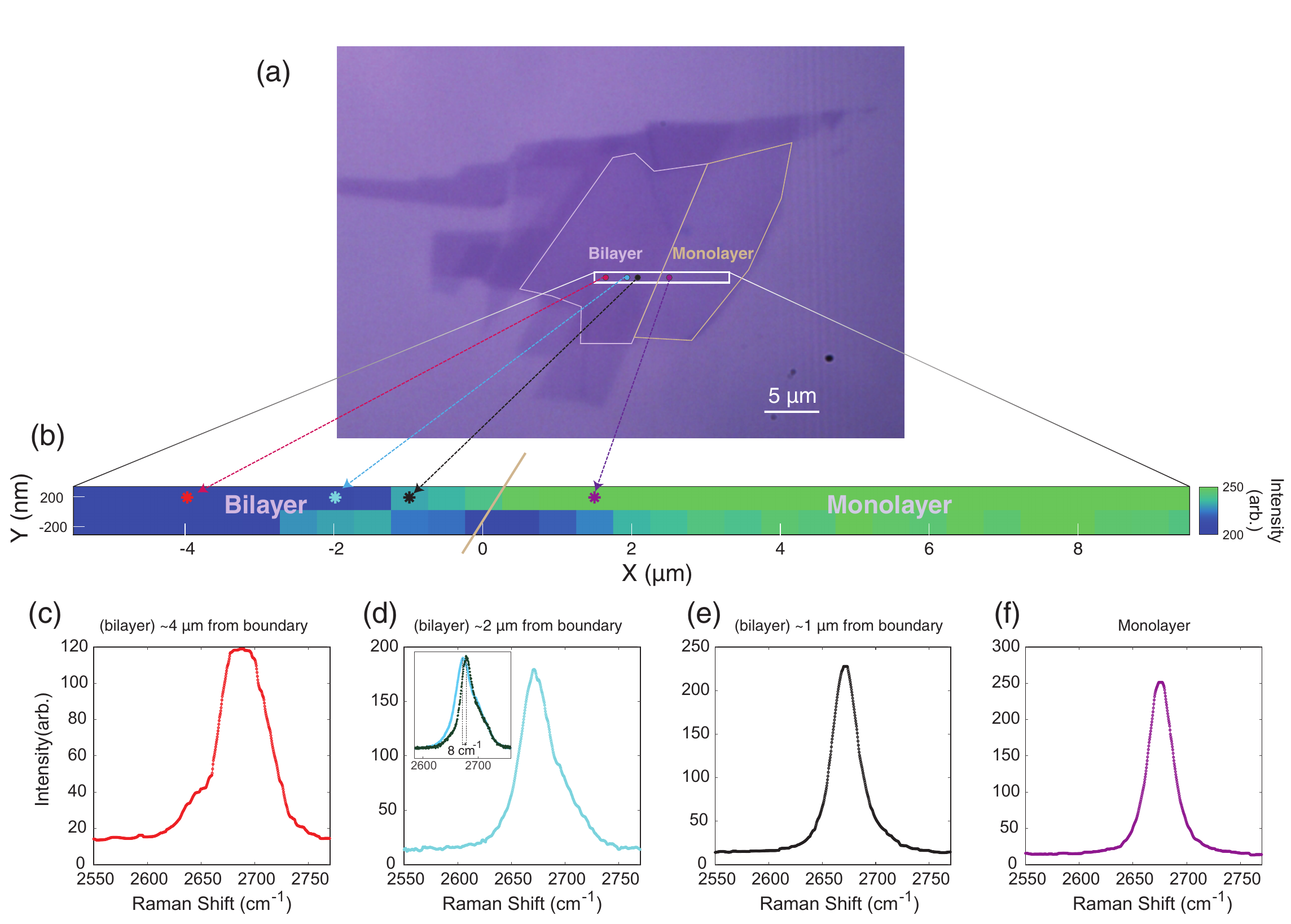}
\caption{\textbf{Raman 2D mode for pristine mono-bilayer graphene sample on SiO\(_2\)}
\textbf{(a)} Optical image of multilayer graphene sample on $285$ nm SiO\(_2\) substrate. Bilayer and monolayer regions are indicated in the image. The white box is where Raman spectra were taken. Red, blue, black and purple points from the image are where the representative spectrum was taken. 
\textbf{(b)} Spatial map of intensity of the 2D mode. The boundary between two areas is indicated as white dotted line. 
\textbf{(c-f)} 2D Raman spectrum from same-color point indicated in (b). The title indicates distance from the bilayer-monolayer boundary. Inset in (d) is comparison of the 2D mode measured in the bilayer region of the substrate (blue) and in the configuration 3 of the MEMS sample with twist angle near 0\(^\circ\) (black). The intensity maximum is blueshifted from the bilayer graphene sample by 8 $\text{cm}^{-1}$.
}
\label{fig:E_grboundary}
\end{figure}

For MEMS-substrate devices, the tBG interface is created by the engaging surface, but the rest of the substrate remains covered by monolayer graphene. Thus, the contact area has a mono-bilayer boundary around its periphery, which will be at most $1-2$ \SI{}{\micro\meter} away from the optically measured point. The graphene edge can effectively act as a defect and affect the Raman spectrum \cite{edgeRaman,edgeferrari}. Therefore, to confirm the lineshape of the Bernal bilayer structure near mono-bilayer boundary, we conducted Raman spectroscopy of pristine graphene with mono-bilayer boundary. Previous studies have analyzed the spectrum based on the relative contributions of the bilayer and monolayer 2D modes \cite{edgeRaman,barbosa}. However, we take a different approach because (1) the boundary affects the spectrum beyond simply separating the two regions, and (2) the peak center shifts as the measurement point moves away from bilayer region to monolayer, making it inaccurate to fit the spectrum as a simple combination of bilayer and monolayer spectra.

2D mode Raman spectra measured at different points in the graphene sample are presented in Fig. \ref{fig:E_grboundary}. Optical image of graphene sample on $285$ nm SiO$_2$ substrate is displayed (Fig. \ref{fig:E_grboundary}a) with bilayer-monolayer boundary indicated. Spatial mapping of the Raman 2D mode shows lower intensity in the monolayer region and higher intensity in the bilayer region (Fig. \ref{fig:E_grboundary}) as expected in bilayer and monolayer graphene under the same measurement conditions \cite{zettlraman, ferrari_1}. Beam spot is around $300$ nm in diameter and the spatial resolution of each point in Fig. \ref{fig:E_grboundary}b is 500 nm. The representative spectrum of AB-stack bilayer graphene 2D mode can be measured in the bulk of the bilayer region (\ref{fig:E_grboundary}c). As the measurement spot approaches to within $3$ \SI{}{\micro\meter} of the boundary, the maximum peak shifts its position and its intensity. At approximately $2$ \SI{}{\micro\meter} from the boundary, the line shape becomes different even inside the bilayer region. The peak center shifts and small peak around $2650\text{cm}^{-1}$ is not as pronounced as in the bulk. The overall line shape is strikingly analogous to Fig. \ref{fig:3}d. Direct comparison with the $-0.6\degree$ data is presented as an inset in Fig. \ref{fig:E_grboundary}d. The small shift of the maximum intensity center can be attributed to thermal annealing that the sample undergoes for polymer cleaning \cite{annealshift2D}. As the beam spot approaches closer to the monolayer region, the line shape becomes almost like a monolayer graphene but with a small shift of the center frequency. When the beam reaches the monolayer region, the intensity of the 2D mode is roughly double the intensity measured within the bilayer. These measurements support the assignment of Bernal bilayer structure in Configuration 3 is indeed a Bernal structure that has a mono-bilayer boundary $\sim 2$ \SI{}{\micro\meter} away. 

\begin{figure}
\centering
\includegraphics[width = 6 in]{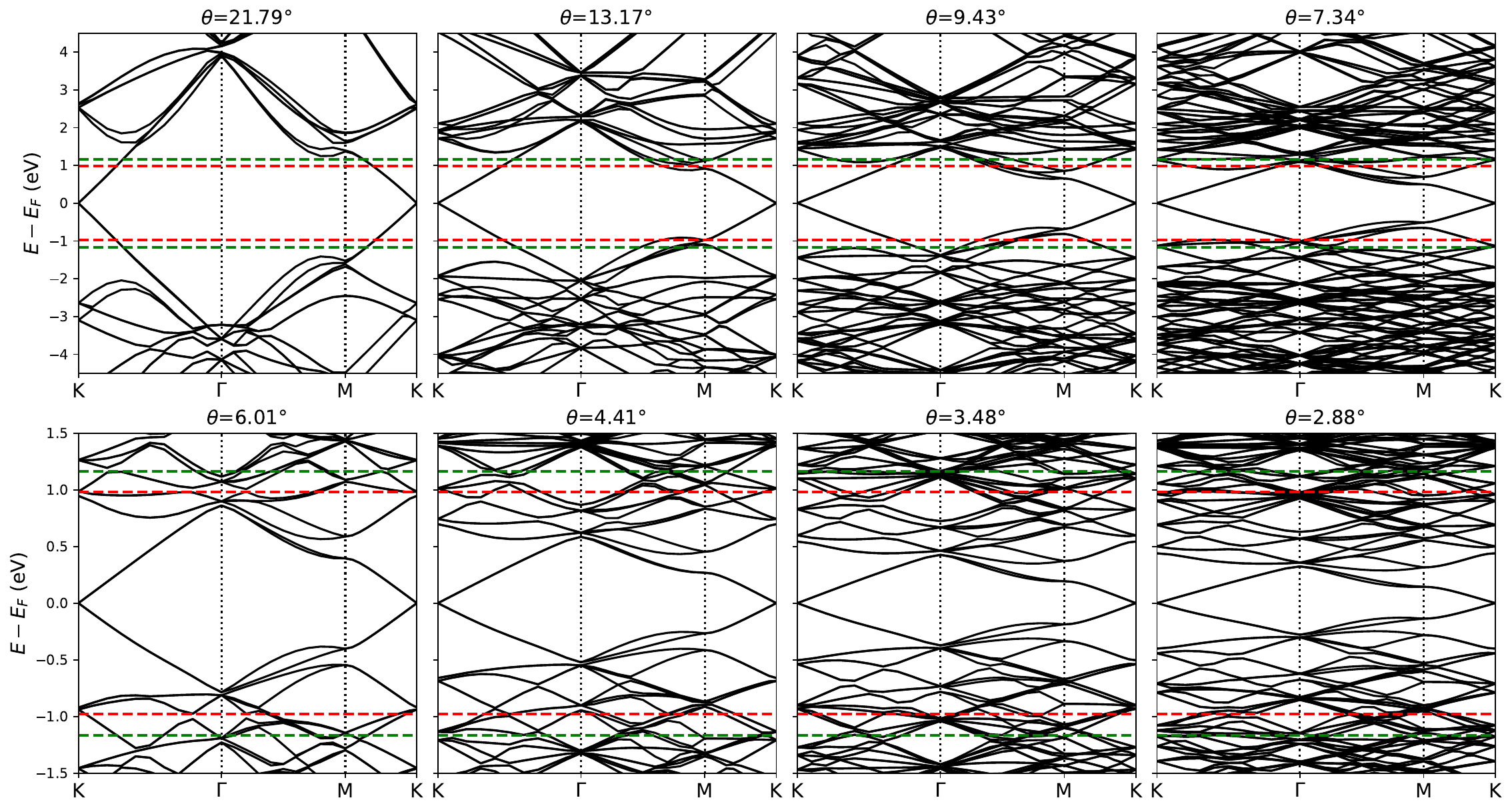}
\caption{\textbf{Twisted bilayer graphene band structure as a function of twist angle, $\theta$. }
The electronic band structure of twisted bilayer graphene calculated using SIESTA DFT for commensurate twist angles in a range relevant for the Raman measurements. The DFT computational framework is described in \cite{DFT_cal}. For the moderate twist angles shown, the main effect of a decreasing twist angle is the reduction of the Fermi velocity, extracted from the slope of the bands at K point. Around \(22^\circ\), the hybridized band around K point comes close to the laser excitation energy, supporting the experiment that shows strong asymmetry as it approaches \(22^\circ\) (Fig. \ref{fig:3}).
}
\label{fig:E_bandstructure}
\end{figure}

\begin{figure}
\centering
\includegraphics[width = 3 in]{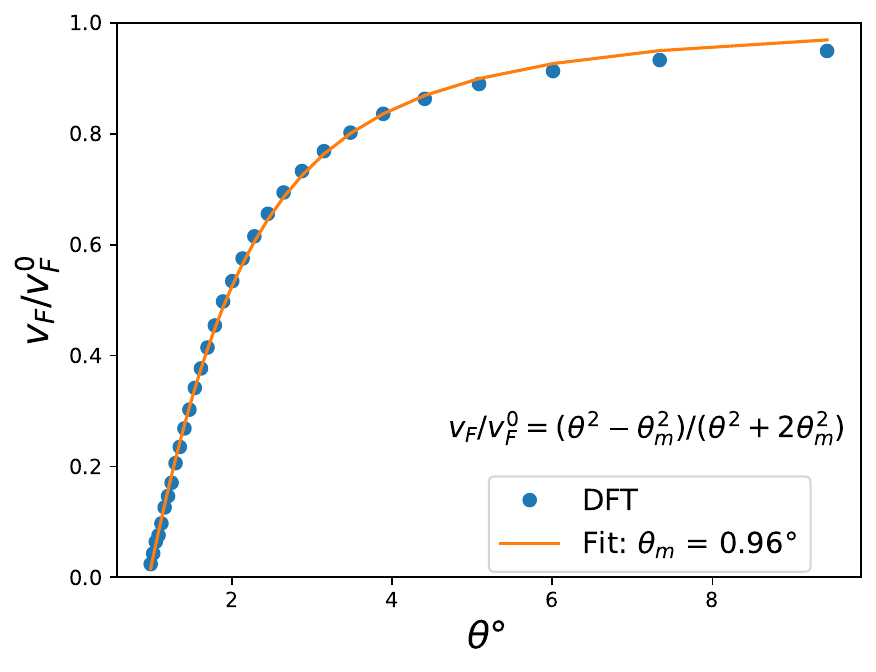}
\caption{\textbf{Extracted Fermi velocity as a function of twist angle}
The Fermi velocity of twist bilayer graphene is extracted from the electronic band structure presented in Fig. \ref{fig:E_bandstructure}, normalized by the Fermi velocity of graphene. The fitting is based on a theoretical framework of \cite{BMmodel} using a single fit parameter \(\theta_m\), which is a 'magic angle' where Fermi velocity vanishes. The fitted \(\theta_m\) is 0.96\(^\circ\).
}
\label{fig:E_vF}
\end{figure}

\section{R peak intensity fit}
\label{SI:Rintensityfit}

In this section, we describe the procedure used to fit the R-peak intensity shown in the main text (Fig.\ref{fig:2}).

A formula based on second-order perturbation theory is used to fit the R-peak intensity shown in Fig. \ref{fig:2}. Following Ref. \cite{dresselhausGband}, the general resonance Raman intensity can be expressed as :
\begin{equation}
    I(E_L, \omega_\text{phonon}) = \sum_i \left| \sum_e \frac{M}{(E_L-E_{ei}-i\gamma)(E_L-E_{ei}-\hbar\omega_\text{phonon}-i\gamma)} \right|^2 \label{eq:SI_int}
\end{equation}
,where $M$ contains the matrix elements for the relevant transitions, including electron-phonon coupling and electron-photon coupling. The subscripts $i$ and $e$ denote the initial and excited states of the electron during the Raman process. $E_L$, $E_{ei}$, and $\omega_\text{phonon}$ are the energy of laser excitation, the energy difference between excited state and initial state, and the phonon energy. $\gamma$ accounts for dissipative broadening. Rewriting the expression in terms of the R peak, Eq. \ref{eq:SI_int} becomes

\begin{equation}
    I(E_L, \omega_\text{phonon}) = \left| \frac{M}{(E_L-E^\text{intra}-i\gamma)(E_L-E^\text{intra}-\hbar\omega_\text{R}-i\gamma)} \right|^2 \label{eq:SI_int_fit_formula}
\end{equation}

where $E^\text{intra}$ denotes intra-cone excitation for the specific angle $\theta$, which can be expressed as Eq. \ref{eq:eLeq},

\begin{equation}
    E_\textbf{intra} (\theta) = \hbar \nu_F \frac{8\pi}{\sqrt{3}a} \sin {\frac{\theta}{2}}
\end{equation}

In Eq. \ref{eq:SI_int_fit_formula}, the excitation energy $E_L$ and the phonon energy $\omega_R$ are known from the measurement. Therefore, the parameters that require fitting are $E_\text{intra}$, or equivalently $\nu_F$ (Fermi velocity at K point), the matrix element $M$, and the dissipative broadening $\gamma$. 

The band structure of tBG is calculated using DFT computational framework to accurately fit the Fermi velocity as a function of twist angle. A few examples of band structures are presented in Fig. \ref{fig:E_bandstructure}. The Fermi velocity of tBG is determined by the slope of the band structure at K point and is normalized by the DFT calculated graphene Fermi velocity ($\nu_0^F$). The ratio is calculated for a wide range of commensurate twist angle and then fit to the functional form for the renormalized velocity derived by truncating the momentum space lattice to the first honeycomb shell (Eq. (11) of \cite{BMmodel}). The fit is based on a single parameter, an effective magic angle $\theta_m$, as the equation shown below.
\begin{equation}
    \frac{\nu_F}{\nu_0^F} = \frac{\theta^2-\theta_m^2}{\theta^2+2\theta_m^2}
\end{equation}
Therefore, the Fermi velocity of tBG can be expressed in terms of the known values (twist angle \(\theta\), effective magic angle \(\theta_m\)) and a normalizing factor \(\nu_0^F\). Therefore, we treat the normalization factor $\nu_0^F$ as a fitting parameter for Eq. \ref{eq:SI_int_fit_formula}. 

The fitted parameters are $ \nu_F^0 = 1.229\pm0.257 \times10^6 $ m/s, $M = 0.3738\pm0.1494$, $\gamma = 0.6217\pm0.1364$ for the $532$ nm measurement and $ \nu_F^0 = 1.227\pm0.916 \times10^6 $ m/s, $M = 0.1589\pm0.2554$, $\gamma = 0.4153\pm0.3647$ for the $633$ nm measurement. The fitting accuracy can be improved by allowing the matrix element to vary with twist angle, as demonstrated using tunable laser at fixed twist angle \cite{CarozoR}.

\section{R peak data in 633 nm measurement}
\label{SI:Rpeak633}

\begin{figure}
\centering
\includegraphics[width = 6 in]{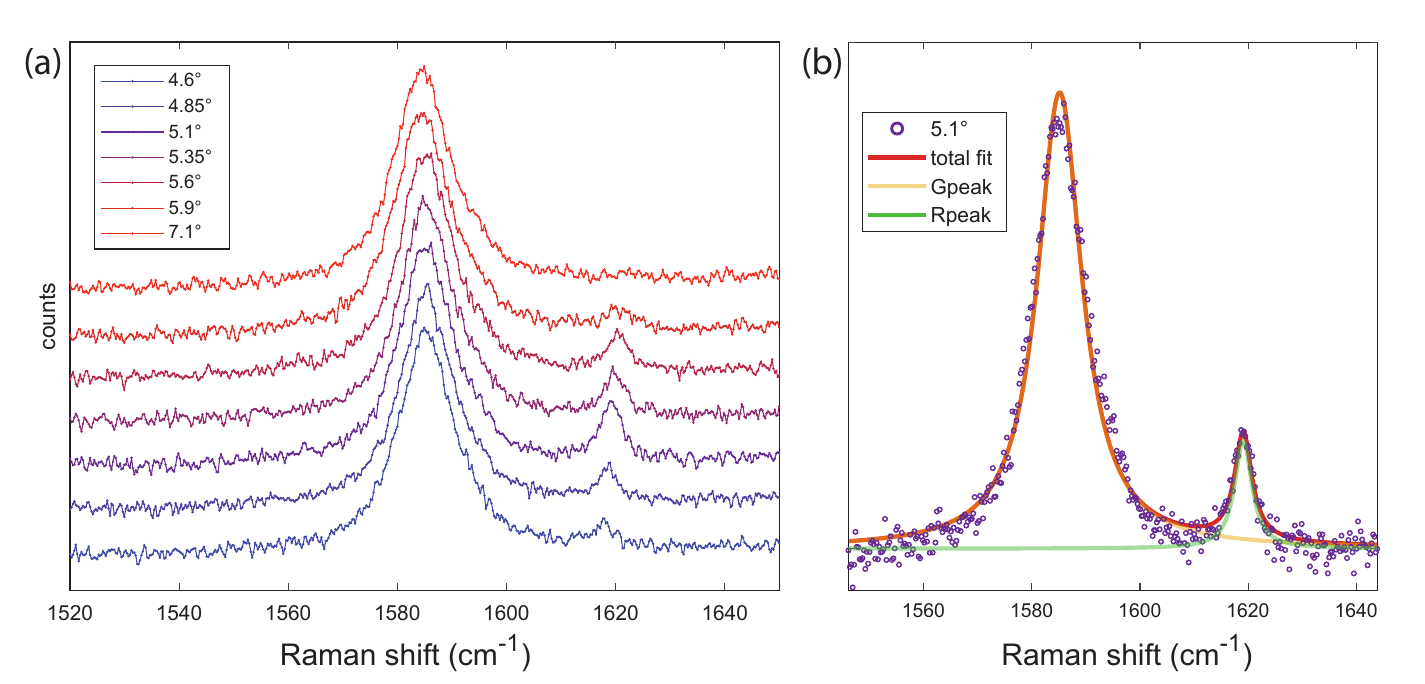}
\caption{\textbf{$633$ nm measurement of `R' peak and fitting}
\textbf{(a)} G mode Raman spectra measured with $633$ nm laser excitation. The small peak around $1620\text{cm}^{-1}$ appears around $5.9\degree$, which persists until $4.6\degree$ twist angle. The fitting results are shown in Fig. \ref{fig:4}d-e. 
\textbf{(b)} Fitted data for $5.1\degree$ rotation. Purple circles are data, yellow and green lines are G and R peak respectively, and the red line is the total fit lines. The G and R peaks are both fitted with Lorentzian functions.
}
\label{figE:Rpeak633}
\end{figure}

To probe a larger angular range for Fig.\ref{fig:4}c, Raman spectroscopy was performed using $633$ nm excitation. The lower excitation energy allows access to phonons with smaller $q_R$ based on Eq. \ref{eq:qtheta} in the main text. Thus, $633$ nm excitation expands the detectable twist angle range by $\sim1.5\degree$. The raw data from the $633$ nm measurement are presented in Fig. \ref{figE:Rpeak633}a. The angle is the absolute twist angle between the two graphene layers. The R peak is fitted using two Lorentzian functions as shown in Fig. \ref{figE:Rpeak633}b. 

\section{Further discussion on theoretical calculations of LO phonon dispersion}
The graphene phonon spectrum was calculated using GW-corrected DFT by \cite{venezuela}. The highest optical branch at $\Gamma$ matches inelastic x-ray scattering (IXS) measurements on graphite quite well. In particular,  \cite{venezuela} note that the DFT frequency of 1561 cm$^{-1}$ for the G peak ($E_{2g}$ $\Gamma$ mode) perfectly fits the IXS measurements. However, that value is 21 cm$^{-1}$ smaller than the measured Raman G peak in graphite (1582 cm$^{-1}$). Such discrepancies have motivated more detailed calculations, including \cite{Erhardt_LOphonon}, which incorporates lattice anharmonicity and non-adiabatic effects such as electron-phonon and phonon-phonon coupling. Furthermore, previous studies have shown how dielectric screening by substrates can affect the electron-phonon coupling and shift Raman peaks \cite{Forster_LORaman_KohnAnomaly}. Given the complexity of the numerical approaches and the sensitivity of the measurements to experimental conditions such as temperature and substrate, we use a simple approach and use the calculated graphene LO spectrum of \cite{venezuela} and apply an overall shift of ~

\clearpage

\end{document}